\documentclass[a4paper,12pt,autodetect-engine]{article}

\usepackage{amsmath}
\usepackage{amssymb}
\usepackage{amsfonts}
\usepackage{amsthm}
\usepackage{subfigure}
\usepackage{graphicx,hyperref,color}
\usepackage[normalem]{ulem} 
\usepackage{cancel} 
\usepackage{cite}
\usepackage{tikz}

\usepackage{subfigure}

\newcommand{\deldel}[2]{\displaystyle{\frac{\partial #1}{\partial #2}}}
\newcommand{\delldell}[2]{\displaystyle{\frac{\partial^{2} #1}{\partial #2^{2}}}}

\newcommand{\del}{\partial}

 \makeatletter
 \@addtoreset{equation}{section}
 \makeatother

\newlength{\subfigwidth}
\newlength{\subfigcolsep}
\newcommand{\e}{\mathrm{e}}
\newcommand{\dop}{\mathrm{d}}
\newfont{\cyr}{wncyr10}
\newcommand{\MyGuillemets}[1]{{\cyr{\symbol{060}}}{#1}{\cyr{\symbol{062}}}}

\date{empty}
\begin{document}
\begin{titlepage}
\null
\begin{flushright}
November, 2025 \\
\mbox{}
\end{flushright}
\vskip 1.7cm
\begin{center}
{\Large \bf 
Quantum Electron Clouds near Black Holes:
\\ \vspace{3mm}
Black Atoms and Molecules
}
\vskip 2cm
\normalsize
\renewcommand\thefootnote{\alph{footnote}}

{\large
Hinako Iseki${}^{1}$\footnote{iseki.hinako(at)st.kitasato-u.ac.jp},
Shin Sasaki${}^{1}$\footnote{shin-s(at)kitasato-u.ac.jp},
and
Kenta Shiozawa${}^{2}$\footnote{shiozawa.kenta(at)kitasato-u.ac.jp}
}
\vskip 0.6cm
  {\it
  ${}^{1}$Department of Physics, Kitasato University \\
  Sagamihara 252-0373, JAPAN
  }
\vskip 0.3cm
  {\it
  ${}^{2}$Center for Natural Sciences, College of Liberal Arts and
  Sciences, \\ 
  Kitasato University \\
  Sagamihara 252-0373, JAPAN
  }
\vskip 1.3cm
\begin{abstract}
We study quantum mechanical wave functions near highly curved spaces,
i.e., black holes.
By utilizing the formalism developed by DeWitt, 
we derive the Schr\"odinger equations in the vicinity of the Schwarzschild 
and the Reissner--Nordstr\"om black hole geometries.
The quantum electron cloud for the ``black hydrogen atom''---an electron
 trapped by black holes---is particularly studied.
We solve the equations and find that 
black holes generally attract the wave functions, localizing them near the
 horizon where the electrons are most likely to be trapped.
These results imply that not only classical objects but also 
the quantum material and even the chemical properties of the atoms are affected by
 strong gravity.
We also discuss black hydrogen molecules composed of multicentered Majumdar--Papapetrou black holes.
\end{abstract}

\end{center}

\end{titlepage}

\newpage
\setcounter{footnote}{0}
\renewcommand\thefootnote{\arabic{footnote}}
\pagenumbering{arabic}
\tableofcontents

\section{Introduction} \label{sect:introduction}

The quantization of gravity remains a difficult issue in modern physics.
Despite many efforts, we have not reached any essential understanding
of quantum theory of gravity.
On the other hand, the relationship between quantum mechanical objects
and gravity is also a subject that deserves investigation.
Regarding this issue, relativistic quantum field theories in curved
spacetimes have been developed \cite{Wald:1995yp}.

Since gravity is the weakest fundamental interaction in nature, it is usually
ignored in quantum theory of elementary particles.
However, it is obvious that in strong gravitational systems, such as
black holes, effects of gravity cannot be ignored.
A typical example is the Hawking radiation of a black hole arising from
the pair production/annihilation of particles in curved spacetimes \cite{Hawking:1974rv, Hawking:1975vcx}.
There are many studies on quantum natures of particles in black hole
backgrounds, for example, see Refs.~\cite{Boulware:1974dm, Hartle:1976tp, Unruh:1976db,
Page:1976df, Parikh:1999mf}.
Very recently, quantum effects in curved spaces have been experimentally
examined \cite{Covey:2025bdj}.

On the other hand, the fate of the quantum matters of intermediate scales
such as $1 \, \text{\AA} = 10^{-10} \, \mathrm{m}$ near black holes has been
the subject of relatively few investigations.
There are several works along this line.
These include one electron atom in general curved spacetimes \cite{Parker:1980hlc, Parker:1980kw}, 
effects of gravity on the hydrogen spectrum \cite{Parker:1982nk}, 
a bound state of relativistic scalar particles and a rotating black hole
\cite{Hod:2015goa}, 
a nonrelativistic limit of scalar and Dirac fields in curved spacetimes
\cite{Falcone:2022hfa, Jentschura:2013bia},
gravitational effects by topological defects on the hydrogen atom
\cite{deAMarques:2002hbv}, and 
general relativistic effects on hydrogen systems
\cite{Fischbach:1981ne}.
There is no doubt that the quantum behavior of matters
 of atomic scale is best studied in the nonrelativistic quantum
 mechanics.
Their quantum mechanical properties, including the detail of the
material, the atomic, and the chemical properties, are completely
understood by the solutions to the Schr\"odinger equation.
It is now tantalizing to study how the strong gravity affects on the quantum
wave functions in the atomic scale.
It is apparent that effects of strong gravity near a black hole cannot
 be captured by simply introducing a Newtonian potential in the
 Schr\"odinger equation.
The Newtonian approximation is valid when a quantum system is far from
the black hole, namely, $r \gg r_{\text{H}}$, where $r_{\text{H}}$ is the
radius of the black hole event horizon.
In order to examine the strong gravitational effects on quantum systems
near the horizon at $r \sim r_{\text{H}}$, where the Newtonian approximation is
violated, the Schr\"odinger equation that incorporates curvature effects
of spacetime is required.

To this end, we utilize the formalism developed by DeWitt \cite{DeWitt:1957at}.
In this formalism, based on the general procedure of the analytical
mechanics in the curved configuration space, the quantum mechanical Schr\"odinger equation
in curved spaces is derived.
Although this formalism subsequently developed into the quantization of
gravity, including the Wheeler--DeWitt equation \cite{DeWitt:1967yk}, 
we use it here to investigate the effects of gravity on quantum
matters of atomic scale.
We note that this nonrelativistic treatment is guaranteed by the local Lorentz
description of the slowly moving condition.
We assume that the local energy and the velocity for particles of mass $m$ obey 
$E_{\rm loc} \ll mc^2$ and $v_{\rm loc}\ll c$.
We also assume that the tidal condition $|R_{\ \nu\rho\sigma}^{\mu}| \,
\lambda^2\ll 1$ is fulfilled 
in the parameter ranges we consider, so that the de Broglie wavelength
$\lambda$ is much shorter than the curvature radius.
This means that the energy scale of quantum particles is smaller than
that of the gravitational tidal forces.
The condition allows us to utilize the nonrelativistic approximation.
Throughout this paper, we focus on black holes of atomic scale.
Then the tidal force condition remains true except within an extremely close region of the
horizon where $|R^{\mu} {}_{\nu \rho \sigma}|^{-\frac{1}{2}}$ and
$\lambda$ are of the same order of magnitude, i.e., the Bohr
radius of the hydrogen.
Therefore, we are interested in the region $r_{\text{H}} < r < \alpha \, r_{\text{H}}$, where
      $\alpha$ is an $\mathcal{O} (10^0 \sim 10^1)$ parameter.
Quantum matters in the vicinity of black holes in other formalisms have
been studied, such as quantum mechanics of constrained particles
\cite{daSilva:2016wxz}, quantum mechanical particles near black holes using Fermi
coordinates \cite{Exirifard:2021cav, Exirifard:2022rsv, Amorim:2023aqw},
and so on.

The organization of this paper is as follows.
In the next section, we briefly introduce the Schr\"odinger equation in
curved spaces that is developed by DeWitt \cite{DeWitt:1957at}.
In Sec.~\ref{sect:Schwarzschild}, we write down the Schr\"odinger
equation in the Schwarzschild black hole background.
We explicitly show the equations in the spherical coordinate that
result from the canonical quantization based on the Cartesian
coordinate.
We decompose the equation by the variable separation and then study the
radial equation by introducing a virtual electric potential.
We show that the Schr\"odinger equation reduces to the confluent Heun's
differential equation, ensuring that it is solved analytically.
We then solve the equation with an appropriate boundary condition and show
that the quantum electron cloud is mostly trapped just on the event horizon.
We visualize the wave functions and find that the probability densities are
strongly affected by the black hole.
In Sec.~\ref{sect:RN_blackhole}, we discuss the charged black holes.
We study the quantum wave functions in the backgrounds of the
nonextremal, extremal, and overextremal Reissner--Nordstr\"om black holes.
The wave functions in the regions inside the inner (Cauchy) and outside the
outer (event) horizons are studied.
We find that the famous gravitational repulsion force near the core of
the Reissner--Nordstr\"om black hole is observed in the wave function.
The extremal limit and the overextremal cases are also discussed.
In Sec.~\ref{sect:blackhole_molecule}, we discuss the black hole
molecule that consists of the two extremal black hole nuclei and one electron.
This corresponds to the hydrogen molecular ion $\mathrm{H}_2^+$.
We show that in a specific region of parameters, the equations can be
solved analytically.
Sec.~\ref{sect:conclusion} is devoted to the conclusion and discussions.
We discuss a possible realization of the black hole atoms in the
present Universe.

\section{Schr\"odinger equation in curved space}
\label{sect:schrodinger}
In this section, we briefly introduce the formalism developed by DeWitt
\cite{DeWitt:1957at}.
A nonrelativistic particle with mass $\mu$ moving in a curved configuration space is
governed by the following Lagrangian:
\begin{align}
L = \frac{\mu}{2} g_{ij} \dot{q}^i \dot{q}^j + a_i \dot{q}^i - V (q),
\end{align}
where $g_{ij}$ is a spatial metric of the configuration space spanned
by the coordinates $q^i$, and $V(q)$ is a potential.
Here, $\dot{q}^i = \frac{{\dop} q^i}{{\dop} t}$, and $t$ is a parameter that specifies the particle motion.
The vector potential $a_i$ should be included when we consider
background magnetic fields.
The Riemann and the Ricci curvature tensors associated with the spatial
metric $g_{ij}$ are defined as
\begin{align}
R^{(3)i} {}_{jkl} 
&=
\del_k \Gamma^i {}_{jl} 
-
\del_l \Gamma^i {}_{jk}
+
\Gamma^i {}_{pk}
\Gamma^p {}_{jl}
-
\Gamma^i {}_{pl}
\Gamma^p {}_{jk},
\notag \\
\Gamma^i {}_{jk} 
&=
\frac{1}{2} g^{ip} 
\Big(
\del_j g_{kp} + \del_k g_{jp} - \del_p g_{jk}
\Big),
\notag \\
R_{ij}^{(3)} &= R^{(3)k} {}_{ikj},
\notag \\
R^{(3)} &= g^{ij} R_{ij}^{(3)},
\end{align}
where $g^{ij}$ is the inverse of $g_{ij}$.

The canonical quantization condition is defined by
\begin{align}
[q^i,q^j] = 0, 
\qquad
[q^i, p_j] = i \hbar \delta^i {}_j,
\qquad
[p_i,p_j] = 0,
\label{eq:cqc}
\end{align}
where $p_i$ is the canonical momentum conjugate to $q^i$.
The quantum Hamiltonian $\hat{H}_{\pm}$ receives the curvature and the metric corrections\footnote{
Note that there is a potential ambiguity on the quantum Hamiltonian.
We have two equivalent choices $\hat{H}_{\pm} = \hat{H} \pm
\frac{\hbar^2}{12} R^{(3)}$ that are consistent with the classical limit $\hbar \to 0$.
We here employ $\hat{H}_+$ rather than $\hat{H}_-$ for definiteness.
This kind of nonuniqueness of quantum theory in curved space is often
recognized. See \cite{Fulling:1972md}, for example.
},
\begin{align}
\hat{H}_{\pm} =& \ \hat{H} \pm \frac{\hbar^2}{12} R^{(3)},
\notag \\
\hat{H} =& \  \frac{1}{2 \mu} g^{-\frac{1}{4}} (p_i - a_i) g^{\frac{1}{2}} g^{ij} (p_j
 - a_j) g^{- \frac{1}{4}} + V,
\label{eq:Hamiltonian}
\end{align}
where $g = \det g_{ij}$.
The coordinate representation for the canonical momentum $p_i$ is given by
\begin{align}
p_i = - i \hbar
\left(
\frac{\del}{\del q^i}
+
\frac{1}{4}
\frac{\del}{\del q^i}
\log g
\right).
\end{align}
In order to ensure the conservation of the probability, 
the $t$-derivative of the wave function $\psi$ is modified in the curved space,
\begin{align}
\frac{\del \psi}{\del t} 
\
\longrightarrow
\
\frac{D \psi}{D t} = \frac{\del \psi}{\del t} + \frac{1}{4}
\left(
\frac{\del}{\del t} \log g 
\right) \psi.
\end{align}
In the end, the Schr\"odinger equation in the curved space is obtained as
\begin{align}
i \hbar \frac{D \psi}{D t} = \hat{H}_+ \psi.
\end{align}
Here, the Hamiltonian $\hat{H}_+$ is given in \eqref{eq:Hamiltonian}.
For later convenience, we rewrite the equation in the explicit form.
In the following, we consider $a_i = 0$ and $q^i$ as the Cartesian coordinate 
\mbox{$x^i$ ($i=1,2,3$)} for a three-dimensional space.
Then, the momentum is given by 
\begin{align}
\hat{p}_i = - i \hbar 
\left(
\hat{A}_i + \frac{1}{4} B_i
\right),
\qquad
\hat{A}_i = \frac{\del}{\del x^i},
\qquad
B_i = \frac{1}{g} \frac{\del g}{\del x^i}.
\end{align}
The curvature-independent part of the Hamiltonian $\hat{H}$ in
\eqref{eq:Hamiltonian} is evaluated as 
\begin{align}
\hat{H} =& \ 
\frac{1}{2 \mu} g^{- \frac{1}{4}} \hat{p}_i g^{\frac{1}{2}}
 g^{ij} \hat{p}_j g^{- \frac{1}{4}} + V
\notag \\
=& \ 
- \frac{\hbar^2}{2 \mu} g^{- \frac{1}{4}}
\Bigg[
\hat{A}_i g^{\frac{1}{2}} g^{ij} \hat{A}_j g^{- \frac{1}{4}}
+
\frac{1}{4} B_i g^{\frac{1}{2}} g^{ij} \hat{A}_j g^{- \frac{1}{4}} 
+
\frac{1}{4} \hat{A}_i g^{\frac{1}{2}} g^{ij} B_j g^{-\frac{1}{4}} 
+
\frac{1}{16} B_i g^{\frac{1}{2}} g^{ij} B_j g^{-\frac{1}{4}}
\Bigg]
+ V.
\end{align}
Note that the derivative $\hat{A}_i$ acts on all parts of its right
side including the wave function.
By distributing the $\hat{A}_i$ to the relevant terms and collecting the
terms altogether, we have the explicit form of the Schr\"odinger equation,
\begin{align}
\scalebox{.96}{$\displaystyle 
i \hbar 
\left(
\frac{\del \psi}{\del t}
+
 \frac{1}{4 g} \frac{\del g}{\del t} \psi
\right)
= 
 -\frac{\hbar^2}{2 \mu} \left\{
\frac{1}{2} g^{ij} \frac{1}{g} \frac{\partial g}{\partial x^i} \frac{\partial \psi}{\partial x^j}
+ \frac{\partial g^{ij}}{\partial x^i} \frac{\partial \psi}{\partial x^j}
+ g^{ij} \left( \frac{\partial}{\partial x^i} \frac{\partial}{\partial x^j} \psi \right)
\right\}
 + 
\left\{
V (x) 
+
\frac{\hbar^2}{12} R^{(3)}
\right\}
\psi.
$}
\label{eq:Seqt}
\end{align}
It is easy to show that the flat space limit $g_{ij} = \delta_{ij}$
results in the ordinary Schr\"odinger equation.

In the following, we focus on the stationary state for $\psi$ and assume
that the background metric $g_{ij}$ is independent of $t$.
In this case, we have $\frac{\del g}{\del t} = 0$, and the wave function
is decomposed as $\psi (t,x) = \e^{-i \frac{E}{\hbar} t} \psi (x)$, 
where $E$ is a constant.
Then, the equation for $\psi (x)$ is found to be
\begin{align}
 -\frac{\hbar^2}{2 \mu} \left\{
\frac{1}{2} g^{ij} \frac{1}{g} \frac{\partial g}{\partial x^i} \frac{\partial}{\partial x^j}
+ \frac{\partial g^{ij}}{\partial x^i} \frac{\partial}{\partial x^j}
+ g^{ij} \left( \frac{\partial}{\partial x^i} \frac{\partial}{\partial
 x^j} \right)
\right\}
\psi (x)
 + 
\left\{
V (x) 
+
\frac{\hbar^2}{12} R^{(3)}
\right\}
\psi (x) = 
E \psi (x).
\label{eq:Seq}
\end{align}

\section{Quantum wavefunction near Schwarzschild black hole}
\label{sect:Schwarzschild}

In this section, we consider the four-dimensional Schwarzschild black hole as the
prototypical example of curved space.
The Schwarzschild solution in the Cartesian coordinate 
is given by
\begin{align}
{\dop}s^2 = 
- \left( 1 - \frac{a}{r} \right) (\dop x^0)^2
+ \sum_{k=1}^3 (\dop x^k)^2
+ \frac{a}{r^2 (r-a)} 
\left(
\sum_{k=1}^3 x^k \, \dop x^k
\right)^2,
\label{eq:Schwarzschild_metric}
\end{align}
where $r^2 = x^2 + y^2 + z^2$, and $a = \frac{2 G_{\mathrm{N}} M}{c^2}$ is a
constant.
A three-dimensional spatial geometry arises from the time 
slice of the metric \eqref{eq:Schwarzschild_metric}.
The point $r = a$ corresponds to the event horizon that the spatial metric
diverges. 
The spatial metric in the Cartesian coordinate in the matrix notation is
given by
\begin{align}
g_{ij} = \left(
\scalebox{.9}{$
\begin{array}{ccc}
1 + \dfrac{a x^2}{r^2 (r - a)} & \dfrac{a x y}{r^2 (r - a)} & \dfrac{a x z}{r^2 (r - a)} \\
\dfrac{a y x}{r^2 (r - a)} & 1 + \dfrac{a y^2}{r^2 (r - a)} & \dfrac{a y z}{r^2 (r - a)} \\
\dfrac{a z x}{r^2 (r - a)} & \dfrac{a z y}{r^2 (r - a)} & 1 + \dfrac{a z^2}{r^2 (r - a)} \\
\end{array}$}
\right),
\quad 
g^{ij} = \left(
\scalebox{.9}{$
\begin{array}{ccc}
1 - \dfrac{a x^2}{r^3} & -\dfrac{a x y}{r^3} & -\dfrac{a x z}{r^3} \\
-\dfrac{a x y}{r^3} & 1 - \dfrac{a y^2}{r^3} & -\dfrac{a y z}{r^3} \\
-\dfrac{a x z}{r^3} & -\dfrac{a y z}{r^3} & 1 - \dfrac{a z^2}{r^3} \\
\end{array}$}
\right).
\label{eq:Sm_Cartesian}
\end{align}
Then we have $g = \det g_{ij} = \frac{r}{r-a}$ and $R^{(3)} = 0$.
We stress that we should begin with the Cartesian coordinate rather than the radial
one since the canonical quantization condition \eqref{eq:cqc} is justified only
in the former coordinate system.
We obtain the correct quantum mechanical equation in the spherical
coordinate $(r,\theta, \phi)$ by starting from \eqref{eq:Schwarzschild_metric} 
in the $(x,y,z)$ system and then switch to the $(r,\theta, \phi)$ system
at the final stage of the calculations.

After tedious calculations, we find that Eq.~\eqref{eq:Seq} 
in the $(r,\theta,\phi)$ coordinate is given by
\begin{align}
- \frac{\hbar^2}{2 \mu}
\Bigg\{
\Big(
1 - \frac{a}{r}
\Big)
\frac{\del^2}{\del r^2}
+
\left(
\frac{2}{r} - \frac{3a}{2r^2}
\right)
\frac{\del}{\del r}
+
\frac{1}{r^2}
\frac{\del^2}{\del \theta^2}
+
\frac{1}{r^2 \sin^2 \theta}
\frac{\del^2}{\del \phi^2}
+
\frac{\cos \theta}{r^2 \sin \theta}
\frac{\del}{\del \theta}
\Bigg\}
\psi
+
V \psi
= 
E \psi.
\label{eq:SBH_radial}
\end{align}
The wave function is decomposed as $\psi (r,\theta,\phi) = R (r) Y_l^m (\theta,\phi)$.
Then, we find that the equations for each part are given by
\begin{align}
&
\left(
1 - \frac{a}{r}
\right) \frac{\del^2 R}{\del r^2}
+
\left(
\frac{2}{r} - \frac{3a}{2r^2}
\right)
\frac{\del R}{\del r}
+
\left\{
\frac{2 \mu}{\hbar^2}
(E - V)
- 
\frac{l (l+1)}{r^2}
\right\} R = 0,
\notag \\
&
\frac{1}{\sin^2 \theta}
\frac{\del^2 Y^m_l}{\del \phi^2}
+
\frac{1}{\sin \theta} \frac{\del}{\del \theta}
\left(
\sin \theta \frac{\del Y^m_l}{\del \theta}
\right)
+
l (l+1) Y^m_l = 0,
\end{align}
where $l = 0, 1, 2, \ldots$ are constants. 
The equation for $Y^m_l$ is the same as the one in the flat space.
Then, $Y_l^m (\theta,\phi)$ is found to be the spherical harmonics 
\begin{align}
&
Y_l^m (\theta, \phi) = 
\sqrt{
\frac{2l+1}{4\pi}
\frac{(l-m)!}{(l+m)!} 
}
P_l^{|m|} (\cos \theta) e^{im \phi}
\notag 
\\
& (l=0,1,2, \ldots,\ m = -l, -l+1, \ldots, l-1, l),
\end{align}
where $P_l^{|m|} (x)$ is the Legendre polynomial.

We next study the equation for the radial direction.
Depending on whether the particles are in a bound state or not, we must
set the boundary conditions appropriately.

\subsection{Free particle near black hole} \label{sect:SB_free}
Before discussing bound states, 
we first consider the case of the free particle $V=0$.
The equation for the radial direction is given by
\begin{align}
\left(
1 - \frac{a}{r}
\right)
\frac{\dop^2 R}{\dop r^2}
+
\left(
\frac{2}{r} - \frac{3a}{2r^2}
\right)
\frac{\dop R}{\dop r}
+
\left(
k^2 - \frac{l (l+1)}{r^2}
\right) R = 0,
\label{eq:free_radial}
\end{align}
where we have assumed $E > 0$ and defined $k^2 = \frac{2 \mu E}{\hbar^2}$.
We can show that Eq.~\eqref{eq:free_radial} has regular singular points at $r = 0$ and
$r = a$.
The indicial equation of \eqref{eq:free_radial} 
reveals that the solutions at $r = a$ behave like $R(r) \sim (r-a)^0$ and $(r -
a)^{\frac{1}{2}}$.
Then, $R(r)$ is regular but $R'(r)$ for $R(r) \sim (r - a)^{\frac{1}{2}}$
diverges at the horizon.
Since the radial coordinate $r$ in \eqref{eq:Schwarzschild_metric} 
becomes timelike in $r < a$ and thus it ceases to be a suitable
coordinate for the wave function, we look for smooth solutions outside the horizon $r > a$.
The Schwarzschild solution is asymptotically flat, and solutions to Eq.~\eqref{eq:free_radial} must approach the ones in the flat space.
It is easy to show that in the asymptotic flat regime $r \to \infty$,
the equation has the solution given by the spherical Bessel functions that behave as
$R(r) \sim \frac{1}{r} \e^{\pm i k r}$.

For consistency, we examine the equation in the asymptotic limit $r \gg a$.
In this region, the equation is compared to that in the flat space.
After rearranging the coefficient of $R''$ to be unity and 
by the rescaling\footnote{
This rescaling arises from matching the integral measure to that of the
flat space $\sqrt{g} |R|^2 \to |R|^2$.
Namely, $R \to g^{- \frac{1}{4}} R = \left( 1 -
\frac{a}{r} \right)^{- \frac{1}{4}} R \sim \left(1 + \frac{a}{4r}\right)
R$.
} 
$\hat{R} = \left(1 + \frac{a}{4r} \right) R
$,
the equation results in the form
\begin{align}
\hat{R}''(r) + 
\frac{2}{r} \hat{R}'(r)
+
\left(
k^2 - \frac{l(l+1)}{r^2} + \frac{a k^2}{r}
\right) \hat{R}(r) = 0,
\end{align}
where we have ignored $\mathcal{O} (r^{-3})$ terms.
This is the equation in the flat space with the effective Newtonian
potential $V \propto - \frac{1}{r}$. 
Therefore, particles sufficiently away from the black hole certainly perceive
the Newtonian gravity.

We now rewrite Eq.~\eqref{eq:free_radial}.
By introducing the dimensionless coordinate $x = \frac{r}{a}$ and 
substituting $R (x) = \e^{\sigma x} H (x)$, Eq.~\eqref{eq:free_radial} becomes the following form:
\begin{align}
H''(x) + 
\left\{
\frac{\frac{3}{2}}{x}
+
\frac{
\frac{1}{2}
}{x-1}
+
2 \sigma
\right\} H'(x)
+
\frac{1}{x (x-1)}
\left\{
(2 \sigma + a^2 k^2) x 
-
\frac{3}{2} \sigma - l (l+1)
\right\} H (x) = 0,
\label{eq:Heun_diff}
\end{align}
where we have set $\sigma = \pm i a k$.
We find that \eqref{eq:Heun_diff} is the 
confluent form of Heun's differential equation.
The parameters\footnote{
The standard form of Heun's differential equation of confluent type is 
$
H''(z) + 
\left[
\frac{\gamma}{z} + \frac{\delta}{z-1} + \varepsilon
\right] H'(z)
+ \frac{\alpha z - q}{z (z-1)} H (z) = 0.
$
} are $\gamma = \frac{3}{2}$, 
$\delta = 
\frac{1}{2}
$,
$\varepsilon = 2 \sigma = \pm 2 i a k$, 
$\alpha = 2 \sigma + a^2 k^2$, and 
$q = - \frac{3}{2} \sigma - l (l+1)$.
The general solution is, therefore, given by
$
R (r) = \e^{\pm i k r} H (r)
$,
where $H(r)$ is the confluent Heun function defined by the parameters given above.
The signs $\pm$ correspond to the outgoing and ingoing modes,
respectively.
The confluent Heun function has two linearly independent asymptotic
expansions at $x \to \infty$: $H (x) \sim x^{-\alpha/\varepsilon}$ and
$H(x) \sim \e^{- \varepsilon x} x^{-\gamma 
- \delta + \alpha/\varepsilon}$ \cite{Ronveaux:1995}.
The physically relevant solution that behaves as $R(r) \sim \frac{1}{r}
\e^{\pm i k r}$ ($r \to \infty$) corresponds to the first branch.
Physically, this implies that a purely outgoing or ingoing mode at infinity originates from the scattering of waves at the horizon.
The local analysis at $r=a$ gives $R \sim A + B\sqrt{r-a}$, where $A,B$ are constants. 
A boundary condition at infinity, which we will take such that $R \sim \frac{\sin k r}{r}$ is a real-valued function, fixes a particular linear combination of these two local behaviors through the global properties of the confluent Heun function, or equivalently, through a numerical integration.

\begin{figure}[t]
\setlength{\subfigwidth}{.99\linewidth}
\addtolength{\subfigwidth}{-.5\subfigcolsep}
\begin{minipage}[b]{\subfigwidth}
\centering
\subfigure[Radial probability density of the $l=1$.]{\includegraphics[width=\textwidth]{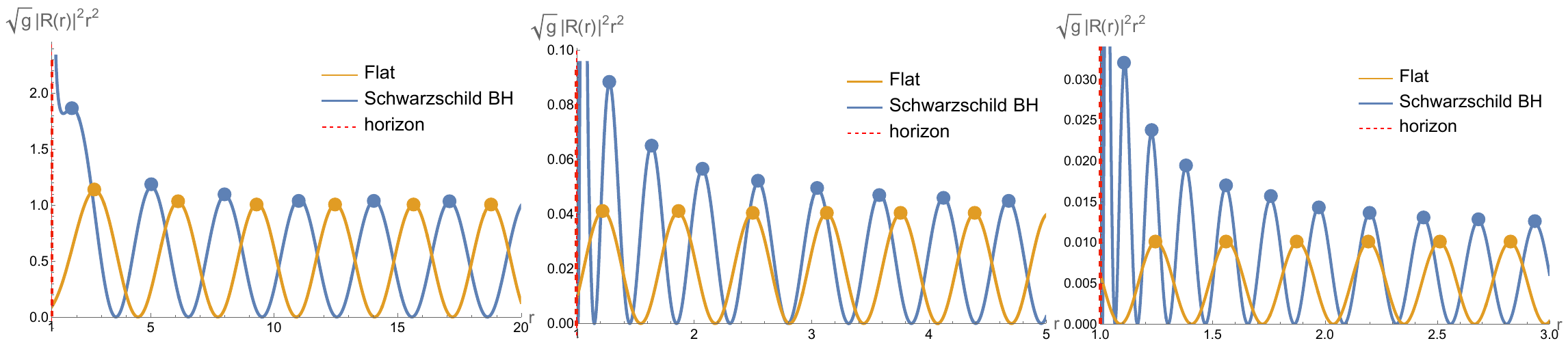}}
   \end{minipage}
\begin{minipage}[b]{\subfigwidth}
\centering
\subfigure[Radial probability density of the $l=2$.]{\includegraphics[width=\textwidth]{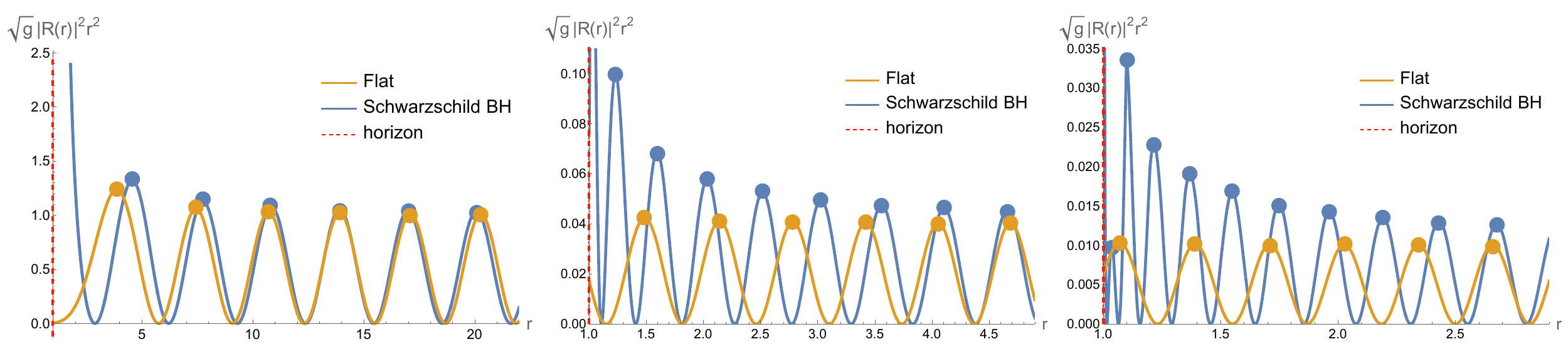}}
\end{minipage}
\caption{%
The probability density $\sqrt{g} |R (r)|^2 r^2$ in the radial
 direction.
Comparison between the flat space and the black hole background cases
for $a=1$. 
(a) $l = 1$ with 
$k=1$ (left),  
$k=5$ (middle), 
and 
$k=10$ (right), and 
(b) $l = 2$ with 
$k=1$ (left),  
$k=5$ (middle), 
and 
$k=10$ (right).
The red dotted line corresponds to the event horizon $r = a$.
The infinity is set to be $r_{\text{max}} = 1000$ and we define the
 boundary condition $R (r_{\text{max}}) = 0$ to find the numerical
 solutions with \textit{Mathematica}.
The normalization factors are appropriately chosen.
}
\label{fig:free_radial}
\end{figure}

In order to see the behavior of the wave function near the black hole, it
is useful to visualize the solution.
To this end, we solve the equation numerically and make plots.
The probability density $\sqrt{g} |R(r)|^2 r^2$ in the radial direction is shown in
Fig.~\ref{fig:free_radial}.

\begin{figure}[t]
\centering
\includegraphics[width=.8\textwidth]{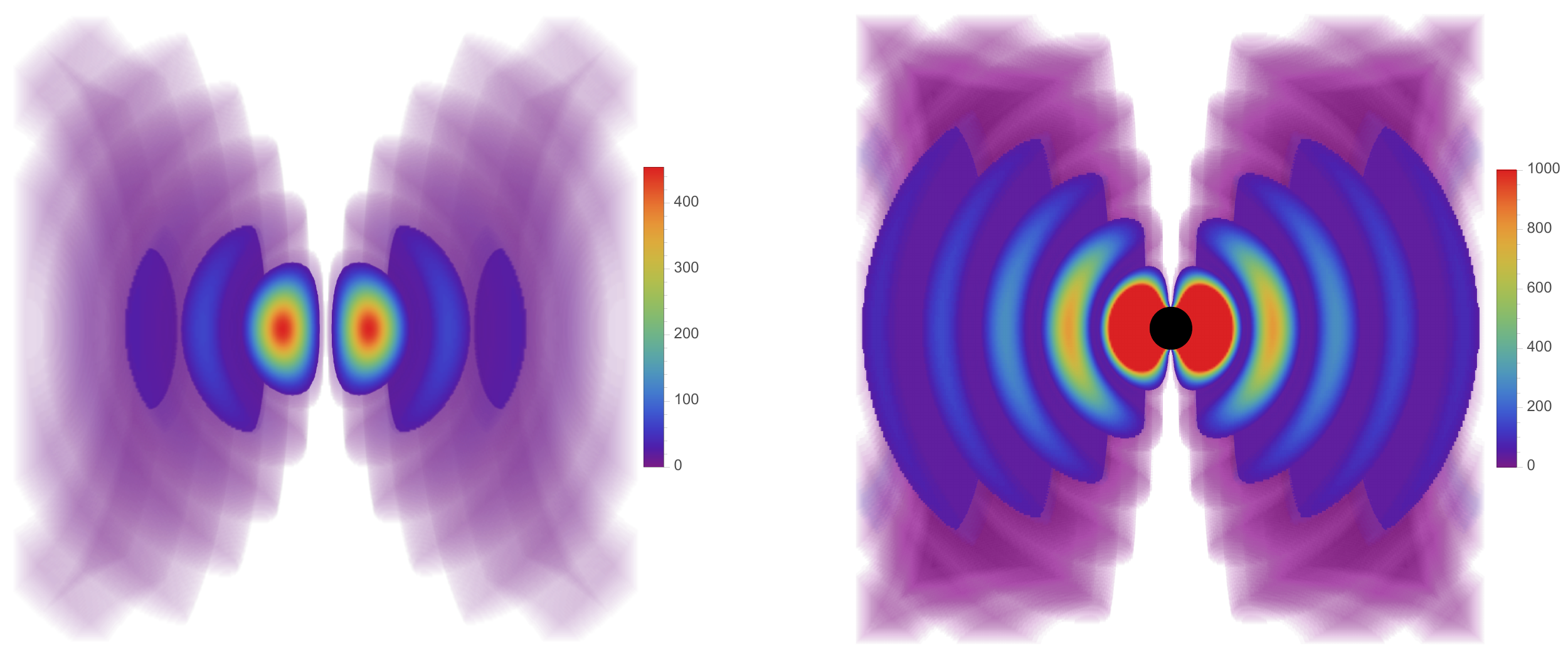}
\caption{%
The 3D plots for the probability density $|\psi (r, \theta, \phi)|^2$ for free particles.
The cases of flat space (left) and the black hole background (right).
The parameters are fixed to $a =1$, $l = 1$, and $k =1$.
The values increase from blue to red and 
the event horizon is shown in the black sphere.
}
\label{fig:free_3d}
\end{figure}

We find that the wave function is most strongly attracted near the
horizon.
At the same time, the wave function exhibits overall shifts toward the
black hole direction.
Note that although $R(r)$ itself is finite at $r = a$, 
the probability density diverges at the horizon due to the factor
$\sqrt{g} = \sqrt{\frac{r}{r-a}}$.
It is also observed that the higher the energy, the more it is attracted
toward the center.
This can be confirmed by comparing the distance between any two peaks in
the wave function.
As the energy increases by $k=1, 5, 10$, the distances between any
adjacent peaks become smaller.
A naive interpretation of this fact is that objects with higher energy
(mass) are attracted more strongly by gravity.
These characteristic behaviors indicate that quantum particles are also
strongly affected by gravity.
The 3D plots in Fig.~\ref{fig:free_3d} clearly show how gravity affects
the wave function.
Note that we employ real-valued solutions for $R(r)$.
This means that the linear combination of the ingoing and outgoing modes
produces the standing wave, resulting in the vanishing of the radial
current of probability $j_r \propto R \del_r \bar{R} - \bar{R} \del_r R
= 0$.
%
%

In the following, we introduce an electric attractive force potential and 
examine a bound state in the black hole background.

\subsection{Attractive force by potential} 

In the next section, we will discuss a genuine 
``black hole hydrogen atom,'' namely, a negatively charged particle trapped by a
positively charged black hole.
Before discussing this truly ideal situation, we first consider a simpler
setup, i.e., a particle subject to a hypothetical attractive
potential in the Schwarzschild geometry.
We stress that the Schwarzschild black hole cannot support a static electric potential in the strict sense.
A charged source would inevitably change the Schwarzschild geometry to the one for the Reissner--Nordstr\"{o}m solution. 
However, before proceeding to the rigorous treatment of the case in the
next section, we first examine this setting as a simplified toy model to
find the effects of the curved background on the hydrogenlike
potential.
This setup is justified by introducing a small probe charge to the
Schwarzschild black hole and ignoring its backreaction to the geometry.
We assume that there is the potential $V = - \frac{e^2}{r}$ producing an
electriclike attractive force. 
We here employ the unit $4 \pi \varepsilon_0 = 1$, and $e$ is the charge
of the particle.
We call this electron in the following.
The equation in the radial direction is given by
\begin{align}
&
\left(
1 - \frac{a}{r}
\right)
\frac{\dop^2 R}{\dop r^2}
+
\left(
\frac{2}{r} - \frac{3a}{2r^2}
\right)
\frac{\dop R}{\dop r}
+
\left\{
\frac{2 \mu}{\hbar^2} 
\Big( E +  \frac{e^2}{r} \Big)
-
\frac{l (l+1)}{r^2}
\right\}
R = 0.\label{eq:sch_pot_radial} 
\end{align}
We find that $r = 0$ and $r = a$ are the regular singular points.
At the horizon $r = a$, the solution behaves like 
$R (r) \sim (r - a)^0$ and $(r - a)^{\frac{1}{2}}$.
Then, $R(r)$ is regular but $R'(r)$ 
for the latter case diverges at $r = a$.

Since the geometry is asymptotically flat, the equation coincide with
the ordinary one for the hydrogen atom in $r \to \infty$.
In this regime the solution should coincides with the well-known
wave function. 
Since the energy spectrum of the hydrogen atom is determined by the
asymptotic boundary condition, 
we can employ the energy spectrum 
$E_n = - \frac{\mu e^4}{2n^2 \hbar^2}$ ($n = 1,2,3, \ldots$)
 for the flat space even in the presence of the black hole.
Following the prescription presented in the previous subsection, we find
that the equation for the radial direction $R(x) = \e^{i \sigma x} H(x)$ is given by
\begin{align}
\scalebox{.99}{$\displaystyle
H''(x) + 
\left\{
2 \sigma + \frac{\frac{3}{2}}{x} 
+ \frac{\frac{1}{2}}{x-1}
\right\}
H'(x)
+
\frac{1}{x (x-1)}
\left\{
(2 \sigma - a^2 \kappa^2 + a s^2) x
-
\frac{3}{2} \sigma
-
l (l+1)
\right\}
H(x) = 0,
$}
\end{align}
where we have again defined 
$x = \frac{r}{a}$, 
$\sigma = \pm  a \kappa$,
$\kappa^2 = \frac{- 2 \mu E_n}{\hbar^2}$, and 
$s^2 = \frac{2\mu e^2}{\hbar^2}$.
The general solution is, therefore, given by the confluent Heun's function
$H(\gamma, \delta, \varepsilon,\alpha,q,x)$, where the parameters are given by 
$\gamma = \frac{3}{2}$, 
$\delta = \frac{1}{2}$, 
$\varepsilon = 2 \sigma$, 
$\alpha = 2 \sigma - a^2 \kappa^2 + a s^2$, and 
$q = - \frac{3}{2} \sigma - l (l+1)$.

\begin{figure}[t]
\centering
\includegraphics[width=\textwidth]{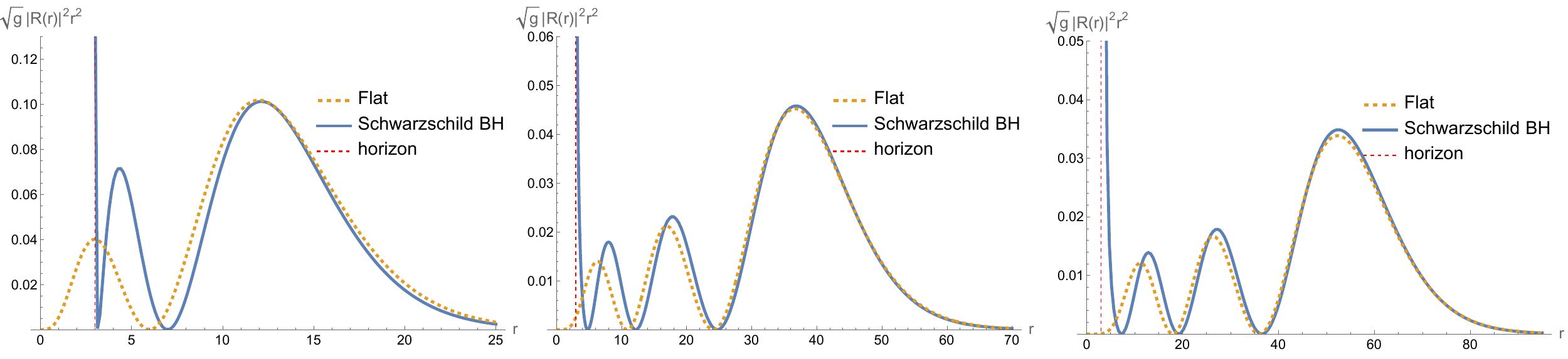}
\caption{%
The radial probability densities of the wave function $\sqrt{g} |R (r)|^2 r^2$.
Comparison between the hydrogen atom in the flat space and in the black hole
 background cases for $a=1$.
The figures correspond to 
the $3p$ orbital ($n = 3$, $l = 1$) (left), 
the $5d$ orbital ($n = 5$, $l = 2$) (middle), 
and the $6f$ orbital ($n = 6$, $l = 3$) (right)
of the hydrogen atom.
We will always use the parameters $\mu = e = \hbar = 1$ in the following.
}
\label{fig:Sch_radial_potential}
\end{figure}

\begin{figure}[t]
\centering
\includegraphics[width=.8\textwidth]{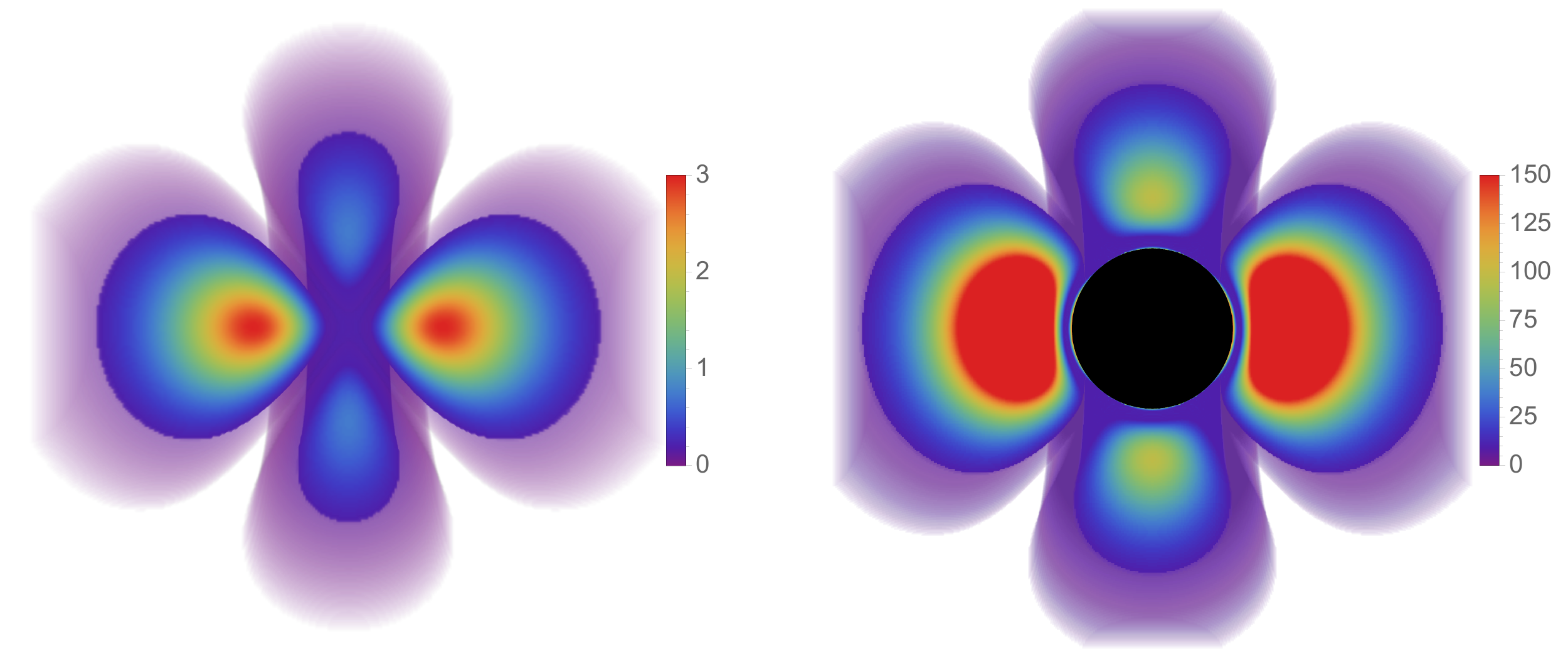}
\caption{%
The 3D plot of the probability density $|\psi|^2$ for the 
particle under the attractive force.
The flat space (left) and the black hole background (right) cases.
The figures correspond to the $3d$ orbital ($n = 3$, $l=2$) of the hydrogen
 atom.
The values increase from blue to red and 
the event horizon $a =5$ is shown in the black color.
}
\label{fig:Sch_3D_potential}
\end{figure}

We find that it is convenient to visualize the wave function by solving
the equation numerically.
To this end, we rewrite Eq.~\eqref{eq:sch_pot_radial} into the following form:
\begin{align}
&\left(\rho-2\kappa a\right)\delldell{u}{\rho}
+\left\{2(l+1+\kappa a)-\rho-\frac{\kappa a}{\rho}(4l+3)\right\}\deldel{u}{\rho}
\notag\\
&\hspace{50mm}
+\left\{
n-l-1-\frac{\kappa a}{2}-\frac{\kappa al}{\rho^2}\left(2l+1\right)+
\frac{\kappa a}{\rho}
\left(\frac{3}{2}+2l\right)
\right\}u=0, 
\label{eq:Sch_pot_eq}
\end{align}
where we have defined $\rho = 2 \kappa r$
and $R(\rho) =\rho^{l} \e^{-\frac{\rho}{2}}u(\rho)$.
We employ the boundary condition $R \to 0 \, (r \to \infty)$.
Then the solutions are shown in Figs.~\ref{fig:Sch_radial_potential}~and~\ref{fig:Sch_3D_potential}. %
%
%
Even though the wave function is strongly concentrated near the horizon,
the peak appears to be displaced to outside 
when it is slightly farther away from the horizon.
This behavior can be interpreted as follows. 
Although electrons are gravitationally attracted toward the black hole 
and extremely localized in a thin shell near the horizon, 
the uncertainty principle causes their kinetic energy to increase when
they are too close to the horizon surface.
Interestingly, the localization of electrons due to the strong
gravitational attraction of the black hole has the opposite effect of
repulsing particles away from it.
These two forces balance to minimize the energy, resulting in a finite spread and peak positions of the wave function.
Paradoxically, the stronger the gravitational attraction toward the center, the farther the electrons are pushed away.
The difference between the bound state and the free particle discussed
in Sec.~\ref{sect:SB_free} is that the former makes standing waves and localizes sharply,
while the latter does not.
Then it is natural that the former is more subject to the uncertainty relation.

\begin{figure}[t]
\setlength{\subfigwidth}{.99\linewidth}
\addtolength{\subfigwidth}{-.5\subfigcolsep}
\begin{minipage}[b]{\subfigwidth}
\centering
\subfigure[Radial probability density of the $3p$ orbital.]{\includegraphics[width=\textwidth]{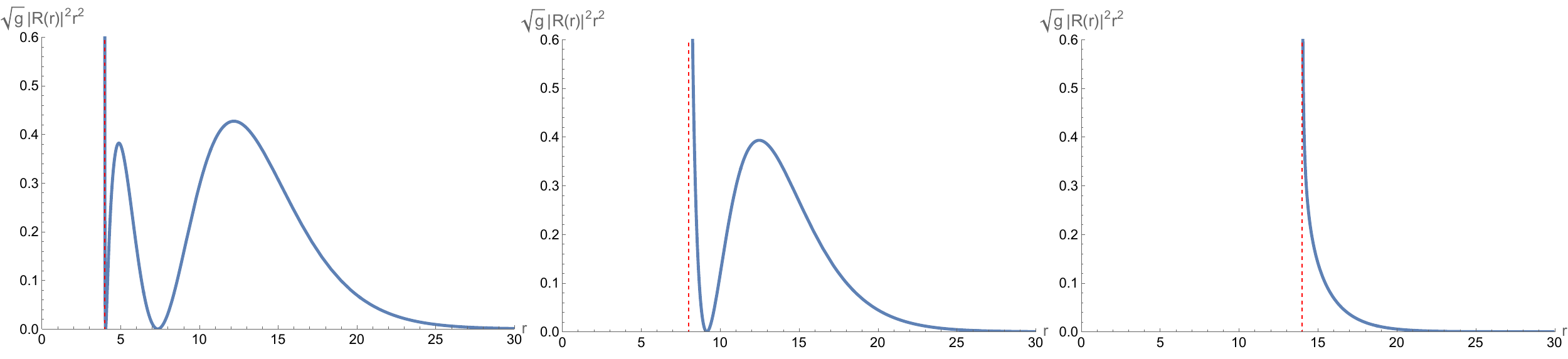}}
   \end{minipage}
\begin{minipage}[b]{\subfigwidth}
\centering
\subfigure[Radial probability density of the $5d$ orbital.]{\includegraphics[width=\textwidth]{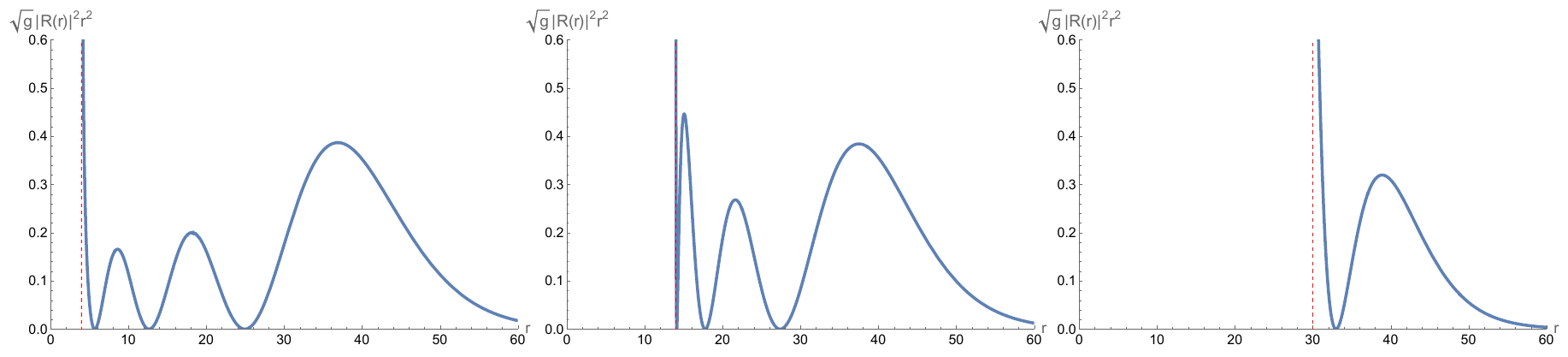}}
\end{minipage}
\caption{%
Comparison of the radial probability density $\sqrt{g}\,|R(r)|^{2} r^{2}$
of (a) the $3p$ orbital ($n=3$, $l=1$) and (b) the $5d$ orbital ($n=5$, $l=2$)  for
 different $a$ values in the Schwarzschild background with the potential $V$.
Each figure shows 
$a = 4$ (left), 
$a = 8$ (middle), and $a = 14$ (right) for the $3p$ orbital,
and $a = 4$ (left), $a=14$ (middle), and $a=30$ (right) for the $5d$ orbital.
The red dotted line denotes the event horizon. 
}
\label{fig:Sch_radial_potential_horizon_sens}
\end{figure}

However, when the black hole becomes sufficiently large, the situation changes.
Figure~\ref{fig:Sch_radial_potential_horizon_sens} shows a comparison of
the radial probability densities obtained by varying the horizon position $r
= a$.
%

%

When the horizon position changes, the peak positions of the
wave function move accordingly.
By changing the horizon position and making it larger than the peak position of 
the probability density, 
the corresponding peak appears to be absorbed into the horizon.
Then gravity eventually prevails over the quantum effects and electrons will
inevitably 
localize on the horizon.

\section{Black hole atoms} \label{sect:RN_blackhole}
We now study the genuine ``black hole atoms.'' 
We introduce the single nonextremal Reissner--Nordstr\"{o}m black hole as the
positively charged atomic nucleus.
In the following, we examine the quantum wave function for a negatively
charged particle trapped by the black hole.
This is nothing but the hydrogen atom composed of a black hole nucleus
and an electron.

\subsection{Black hole hydrogen atom}

The nonextremal Reissner--Nordstr\"om black hole metric in the Cartesian coordinate is
given by
\begin{align}
{\dop}s^2 = 
- \left( 1 - \frac{a}{r} + \frac{Q^2}{r^2} \right) (\dop x^0)^2
+ 
\sum_{k=1}^3 (\dop x^k)^2
+
\frac{
r a - Q^2
}
{
r^2 
\Big(
r^2 - r a + Q^2
\Big)
}
\left(
\sum_{k=1}^3 x^k \, \dop x^k
\right)^2,
\label{eq:RN}
\end{align}
where $r^2 = x^2 + y^2 + z^2$, and $a = \frac{2 G_{\mathrm{N}} M}{c^2}$ and $Q^2 =
\frac{G_{\mathrm{N}} e^2}{c^4}$ are constants. 
The outer and the inner horizons are at $r = r_{+}$ and $r =
r_-$, respectively, where we have defined $r_{\pm} = \frac{a \pm \sqrt{a^2 - 4 Q^2}}{2}$.
The gauge field $A_0 = - \frac{1}{c} \frac{e}{r}$ in the solution provides the electrostatic potential.
The time slice of the solution \eqref{eq:RN} gives a three-dimensional spatial curved
space. The spatial metric is given by
\begin{align}
&
g_{ij} = 
\left(
\begin{array}{ccc}
1 + x^2 f & x y f & x z f \\
x y f & 1 + y^2 f & y z f \\
x z f & y z f & 1 + z^2 f
\end{array}
\right),
\notag \\
&
g^{ij} = 
\frac{1}{1 + r^2 f}
\left(
\begin{array}{ccc}
1 + (y^2 + z^2) f
&
- x y f
&
- x z f
\\
- x y f
&
1 + (x^2 + z^2) f
&
- y z f
\\
- x z f
&
- y z f
&
1 + (x^2 + y^2) f
\end{array}
\right),
\notag \\
&
f = \frac{r a - Q^2}{r^2 (r^2 - r a + Q^2)}.
\end{align}
The Ricci scalar of the three-dimensional space is calculated as
\begin{align}
R^{(3)} = \frac{2 Q^2}{r^4}.
\end{align}
We also have $g = \det g_{ij} = \frac{r^2}{(r - r_+) (r - r_-)}$.
Then, after some calculations, the Schr\"odinger equation in the
$(r,\theta,\phi)$ system is given by
\begin{align}
&
- \frac{\hbar^2}{2 \mu}
\Bigg\{
\left(
1 - \frac{a r - Q^2}{r^2}
\right)
\frac{\del^2}{\del r^2}
+
\left(
\frac{2}{r} - \frac{3 a r - 2 Q^2}{2 r^3}
\right)
\frac{\del}{\del r}
+
\frac{1}{r^2}
\frac{\del^2}{\del \theta^2}
+
\frac{\cos \theta}{r^2 \sin \theta} 
\frac{\del}{\del \theta}
+
\frac{1}{r^2 \sin^2 \theta}
\frac{\del^2}{\del \phi^2}
\Bigg\} \psi
\notag \\
& \qquad
+
\frac{\hbar^2 Q^2}{6 r^4} \psi 
-
\frac{e^2}{r} \psi = E \psi.
\end{align}
Note that the Coulomb potential $V = - \frac{e^2}{r}$ that gives
the electromagnetic attractive force
is naturally incorporated through the coupling $e c A_0 \psi$ in the
Schr\"odinger equation.
This differs from the case of the Schwarzschild black hole where we
have introduced a hypothetical potential by hand.

\begin{figure}[t]
\setlength{\subfigwidth}{.99\linewidth}
\addtolength{\subfigwidth}{-.5\subfigcolsep}
\begin{minipage}[b]{\subfigwidth}
\centering
\subfigure[Radial probability density of the $3p$ orbital.]{\includegraphics[width=\textwidth]{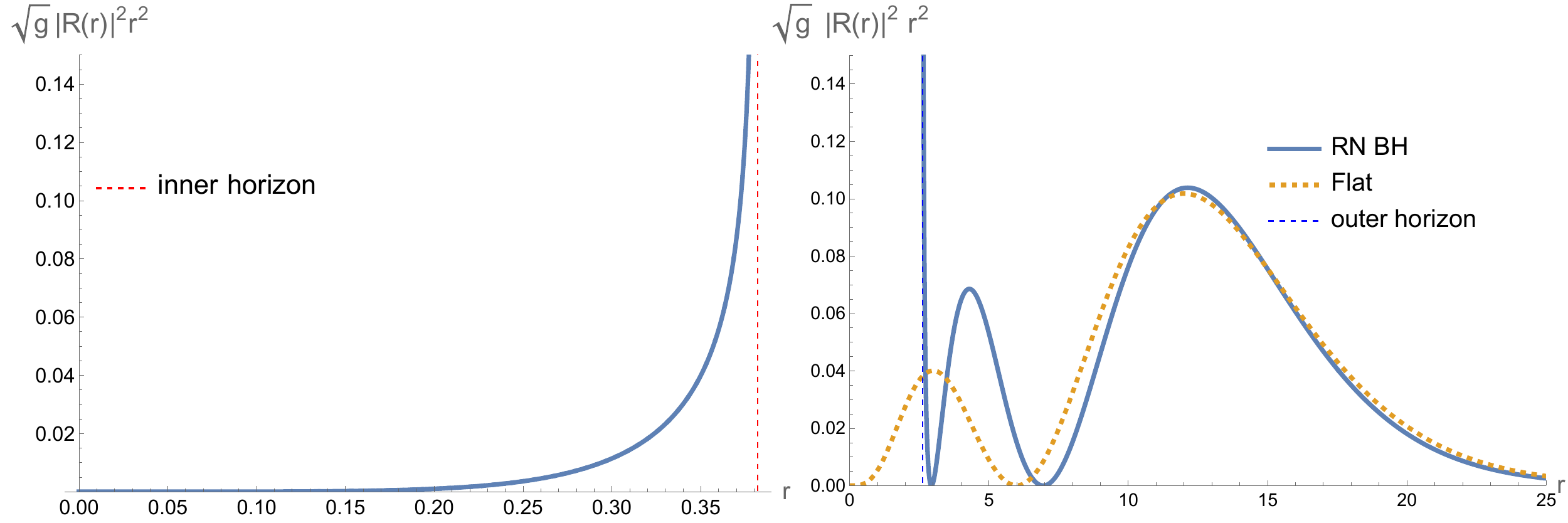}}
   \end{minipage}
\begin{minipage}[b]{\subfigwidth}
\centering
\subfigure[Radial probability density of the $5d$ orbital.]{\includegraphics[width=\textwidth]{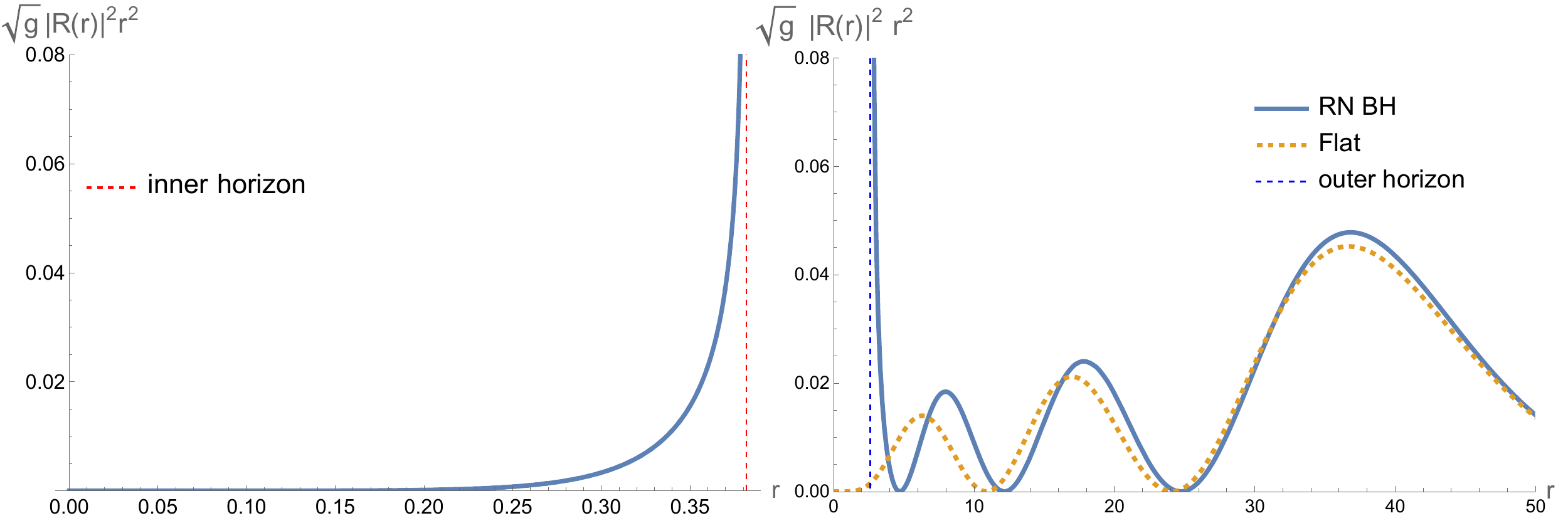}}
\end{minipage}
\caption{%
The probability density $\sqrt{g} |R (r)|^2 r^2$ in the radial direction.
Comparison between the flat space and the Reissner--Nordstr\"om black
 hole 
(RN BH)
background for $a=3$ and $Q=1$.
Inside the inner horizon (left) and outside the outer horizon (right).
The figure corresponds to (a) the $3p$ orbital ($n=3$, $l=1$) 
and (b) the $5d$ orbital ($n=5$, $l=2$)
of the hydrogen atom.
}
\label{fig:charged_radial}
\end{figure}

By substituting the decomposition $\psi (r, \theta, \phi) = R (r) Y
(\theta, \phi)$ in the equation, we find that the angular function 
$Y(\theta, \phi)$ satisfies the same equation in the Schwarzschild case,
and hence it is the spherical harmonics.
On the other hand, the equation in the radial direction is given by
\begin{align}
\left(1-\frac{a r - Q^{2}}{r^{2}}\right)\frac{\partial^{2}R}{\partial r^{2}}
+\left(\frac{2}{r}-\frac{3 a r - 2 Q^{2}}{2 r^{3}}\right)\frac{\partial R}{\partial r}
+\left\{\frac{Q^{2}}{3 r^{4}}+\frac{2 \mu}{\hbar^{2}}\left(E+\frac{e^{2}}{r}\right)
-\frac{l(l+1)}{r^{2}}\right\}R=0.
\label{eq:RN_radial}
\end{align}
It is shown that $r = 0$ and $r = r_{\pm}$ are regular singular points.
At $r \sim 0$, the solution behaves like
\begin{align}
R(r) \underset{r \to 0}{\sim} 
\exp 
\Bigg[
i \sqrt{\frac{1}{3Q}} \ln r
\Bigg],
\qquad
R'(r) \underset{r \to 0}{\sim} 
\frac{1}{r}
\exp 
\Bigg[
i \sqrt{\frac{1}{3Q}} \ln r
\Bigg].
\label{eq:RN_r0}
\end{align}
Therefore, $R(r)$ is regular but $R'(r)$ diverges at $r = 0$.
Near the horizons $r \sim r_{\pm}$, the solution behaves like
\begin{align}
R (r) \underset{r \to r_{\pm}}{\sim} (r - r_{\pm})^0, \ (r -
 r_{\pm})^{\frac{1}{2}}.
\label{eq:RN_regularity_horizon}
\end{align}
Then, $R(r)$ is regular at the horizons, but 
for $R(r) \sim (r - r_{\pm})^{\frac{1}{2}}$
its derivative
diverges, as in the case of the Schwarzschild black hole.
Now we look for bound states where $E < 0$.
The asymptotic consistency again results in 
$E_n = - \frac{\mu e^4}{2n^2 \hbar^2}$.
Using the same parameters as discussed before, 
Eq.~\eqref{eq:RN_radial} is rewritten as
\begin{align}
& \scalebox{.95}{$
\begin{aligned} \displaystyle
&
(\rho^{2}-2\kappa a\,\rho+4\kappa^{2}Q^{2})\,\frac{\partial^{2}u}{\partial\rho^{2}}
+\Bigl\{-\rho^{2}+2\rho\,(l+1+\kappa a)-\kappa a\,(4l+3)
+\frac{4\kappa^{2}Q^{2}}{\rho}\,(2l+1-\rho)\Bigr\}\frac{\partial
 u}{\partial\rho}
\notag \\
&
+\Bigg[\rho\Bigl(n-l-1-\frac{\kappa a}{2}\Bigr)
+\kappa a\Bigl(2l+\frac{3}{2}\Bigr)
-
\left\{
\frac{\kappa a\,l}{\rho}
+\frac{2\kappa^{2}Q^{2}}{\rho}
\right\}
(2l+1)
+\frac{4\kappa^{2}Q^{2}}{\rho^{2}}\Bigl(l^{2}+\frac{1}{3}\Bigr)
+\kappa^{2}Q^{2}\Bigg]\,u=0,
\end{aligned}
$} \notag \\
& 
\label{eq:RN_u}
\end{align}
where $\rho = 2 \kappa r$ and $R (r) = \rho^l \e^{- \frac{\rho}{2}} u
(\rho)$ as in the case of Eq.~\eqref{eq:Sch_pot_eq}.
We solve this equation numerically.
The numerical visualization of the wave functions is shown in
Fig.~\ref{fig:charged_radial}. 

The region outside the outer horizon behaves similarly to the case of
the Schwarzschild black hole.
The wave function is strongly attracted to the outer horizon.
On the other hand, we find that within the region inside the inner horizon, the wave function is attached
to the inner horizon from the interior. 
This behavior is most naturally interpreted as the effect of the gravitational
repulsion force observed near the core of the Reissner--Nordstr\"om
black \mbox{hole~\cite{Mahajan:1979dg, Qadir:1983}}.
This repulsive force is intuitively understood from the fact that $g_{00}
\sim - \frac{a}{r} + \frac{Q^2}{r^2}$ in~\eqref{eq:RN} can be identified
with the gravitational potential in the weak gravity approximation.
In the region $r \gg a$, the attractive force from $- \frac{a}{r}$ is
dominant, while in the small $r$ region, the repulsive force from
$\frac{Q^2}{r^2}$ overcomes the former.

\begin{figure}[t]
\centering
\includegraphics[width=.999\textwidth]{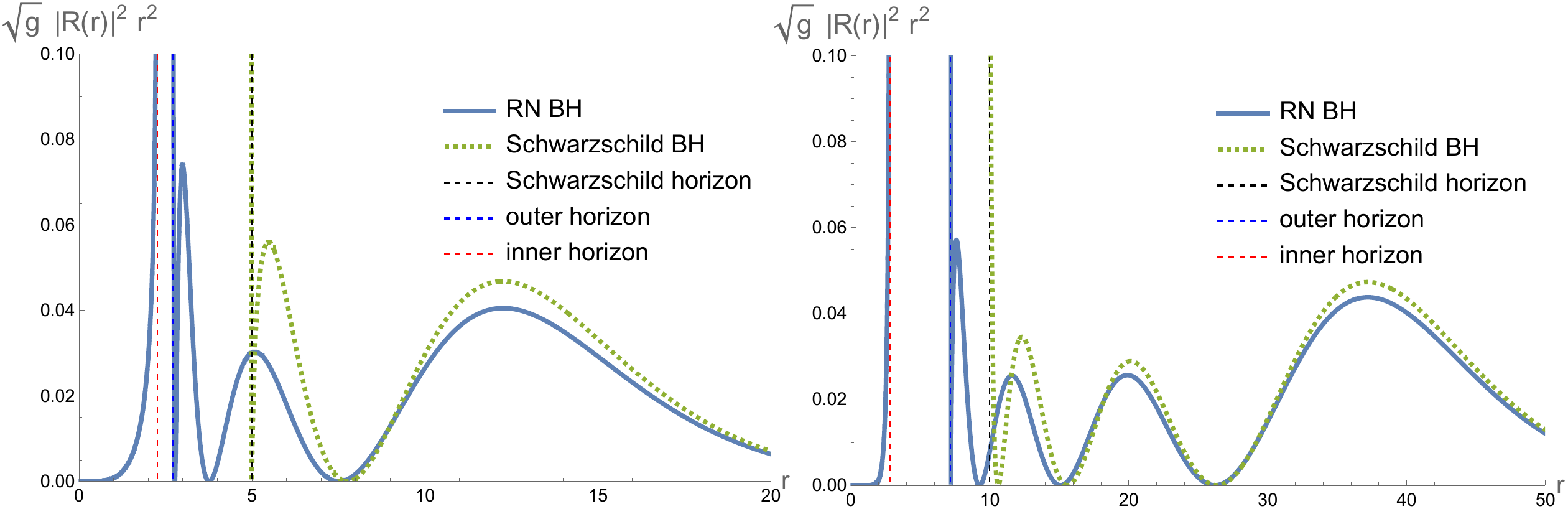}
\caption{%
Comparison of the radial probability density in the Schwarzschild black hole
 and the Reissner--Nordstr\"om black hole backgrounds.
The left figure shows the $3p$ orbital ($n=3,l=1$) with the parameters 
fixed at $a=5$ and $Q=2.49$, while the right figure shows the $5d$ orbital 
($n=5,l=2$) with the parameters fixed at $a=10$ and $Q=4.5$.
In the Reissner--Nordstr\"om case, new peaks appear near the outer horizon, 
indicating the characteristic influence of the electromagnetic charge on the 
radial distribution.
There appear new peaks close to the outer horizon in the Reissner--Nordstr\"om case.
}
\label{fig:flat_sch_RN}
\end{figure}

The radial probability densities for the Schwarzschild and the Reissner--Nordstr\"om black
holes are shown in Fig.~\ref{fig:flat_sch_RN}. 
We observe that there appear new peaks in the Reissner--Nordstr\"om
black hole background which are absent in the flat and the Schwarzschild cases.
We can see that the radial wave function $R(r)$ oscillates more rapidly
when it closes to $r = r_+$.
This indicates that there appear more stable orbits of electrons in the
vicinity of horizons and we expect that the chemical property of the black hydrogen atom is quite
different from the ordinary one.

\subsection{Extremal limit}
In the extremal limit $a^2 = 4 Q^2$, the two horizons coincide, $r_+ = r_-$.
The behavior of the solution~\eqref{eq:RN_regularity_horizon} is
inherited to this case.
As we approach the extremum $4 Q^2 \to a^2$, 
the distance between the inner and the outer horizons decreases 
and it becomes eventually zero at $4Q^2 = a^2$.
In this limit, the gravitational repulsion takes place just inside the
degenerate horizon.

\begin{figure}[t]
\setlength{\subfigwidth}{.99\linewidth}
\addtolength{\subfigwidth}{-.5\subfigcolsep}
\begin{minipage}[b]{\subfigwidth}
\centering
\subfigure[Radial probability density of the $3p$ orbital.]{\includegraphics[width=\textwidth]{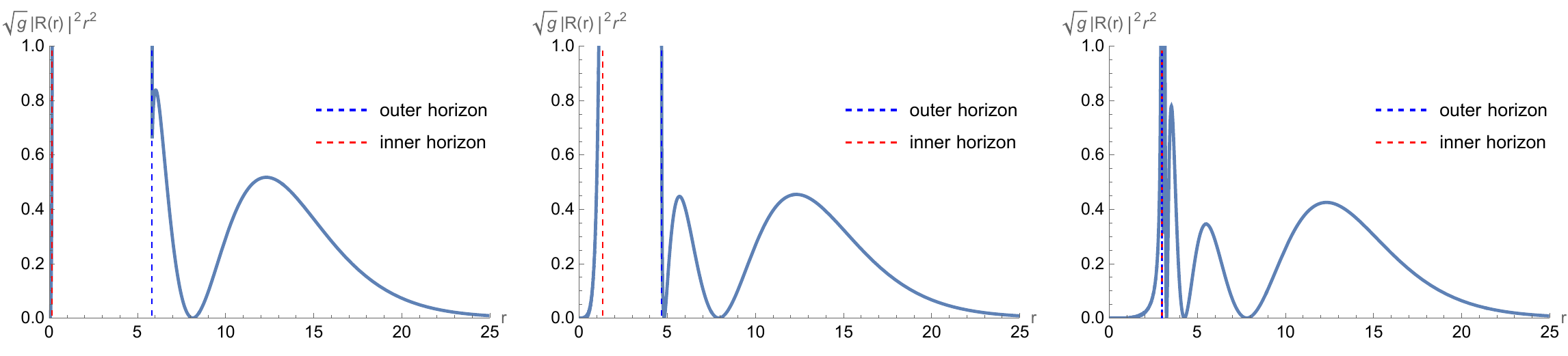}}
   \end{minipage}
\begin{minipage}[b]{\subfigwidth}
\centering
\subfigure[Radial probability density of the $5d$ orbital.]{\includegraphics[width=\textwidth]{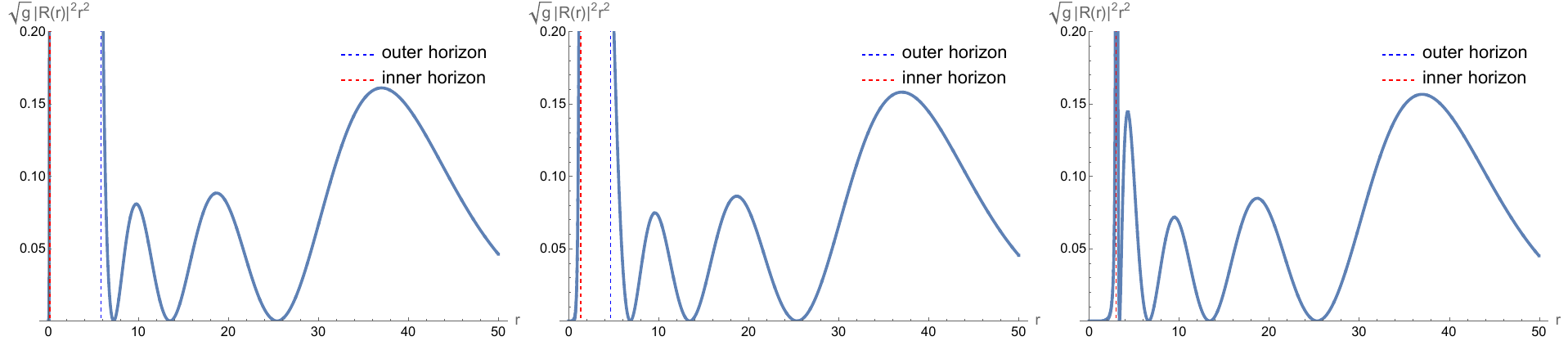}}
\end{minipage}
\caption{%
The radial probability density $\sqrt{g} |R(r)|^2 r^2$
of (a) the $3p$ orbital ($n=3$, $l=1$) 
and (b) the $5d$ orbital ($n=5$, $l=2$)
in the Reissner--Nordstr\"{o}m case.
We fix $a=6$ and vary $Q$ toward the extremal limit $Q=3$.
Each figure shows $Q=1$ (left), $Q=2.5$ (middle), and $Q=3$ (right).
}
\label{fig:extremal_charged_radial}
\end{figure}

\begin{figure}[t]
\centering
\includegraphics[width=.8\textwidth]{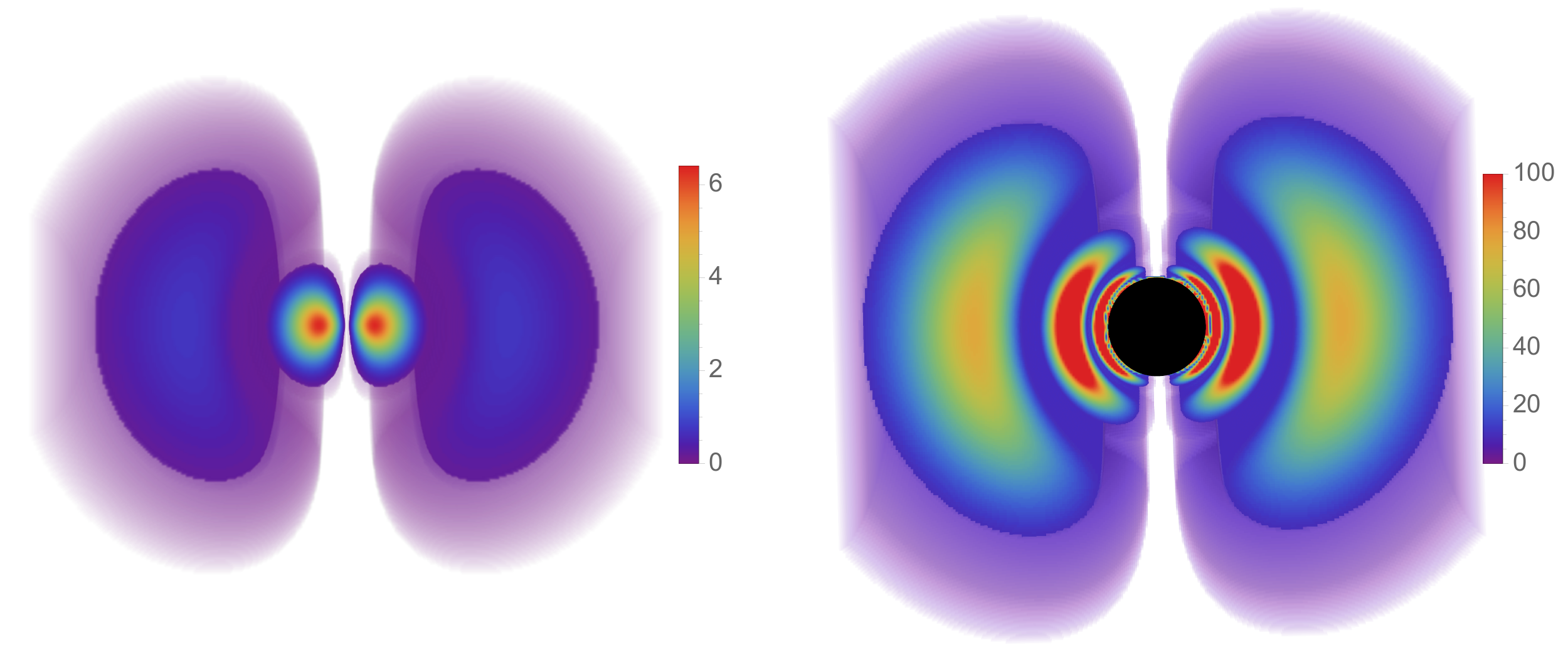}
\caption{%
The 3D plots of the probability density $|\psi|^2$ for the hydrogen atom
 in the $3p$ orbital.
The flat (left) and the extremal black hole backgrounds (right).
The parameters are fixed to $a = 6$ and $Q=3$.
The values increase from blue to red and 
the event horizon is shown in the black color.
}
\label{fig:extremal_3D}
\end{figure}

Figure~\ref{fig:extremal_charged_radial} shows the radial probability
density.
Here, the ratio of the charges between the black hole and the electron is
fixed.
As $Q^2$ increases, the position of the outer horizon moves inward. 
As the horizon shifts toward the center, the peaks of the wave function
that were previously ``eaten away'' reappear. 
This is because increasing $Q$ with fixed $a$ results in a smaller
horizon. 
We can refer to the previous discussion in the Schwarzschild background.
At the same time, as the black hole approaches the extremum, the wave function approaches the horizon
from both inside and outside and it forms a shell of the high probability
density at the degenerated horizon.
This is an electron configuration unique to the black hole atom, not
typically observed in the ordinary hydrogen atom.
The 3D plots of the probability density $|\psi|^2$ are shown in Fig.~\ref{fig:extremal_3D}.


%

\subsection{Overextremal case}
\begin{figure}[t]
\centering
\includegraphics[width=.99\textwidth]{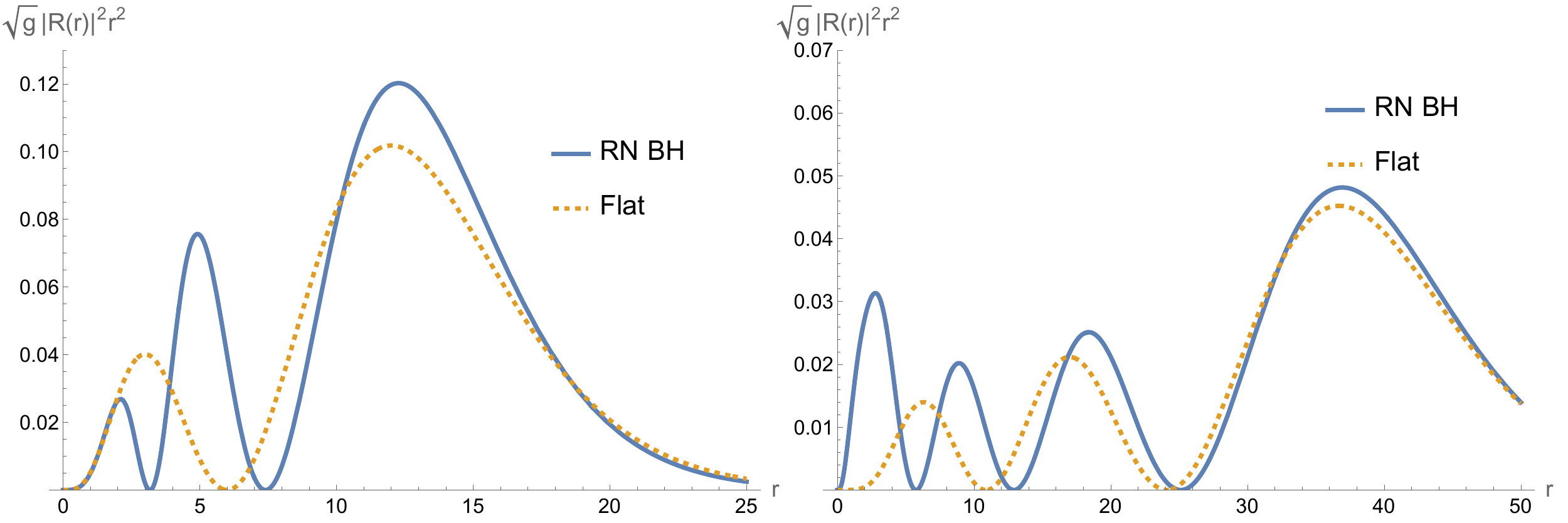}
\caption{%
Comparison of the numerical solution of the radial probability density $\sqrt{g} |R (r)|^2
r^2$ in the overextremal Reissner--Nordstr\"om background and in the flat space.
The parameters are fixed to
$a=5$ and $Q=3$.
The point $r=0$ represents the naked singularity.
The figure on the left corresponds to the $3p$ orbital ($n = 3$, $l = 1$) 
of the hydrogen atom and that on the right to the 
$5d$ orbital ($n = 5$, $l = 2$).
}
\label{fig:overextremal_charged_radial}
\end{figure}

Finally, we examine the overextremal case $4Q^2 > a^2$.
In this case, there are no horizons and it appears naked singularity at $r = 0$.
Even though there is the naked singularity, the solution $R(r)$ is
regular at $r = 0$.
Now the repulsive term $\frac{Q^2}{r^2}$ strongly dominates over the attractive
one $-\frac{a}{r}$ in the overextremal case $Q^2 \gg a^2$ in 
the effective potential $V \sim - \frac{a}{r} + \frac{Q^2}{r^2}$.
This, therefore, should push the wave function further outward from the
naked singularity.
The radial probability density in the overextremal case is shown in
Fig.~\ref{fig:overextremal_charged_radial}.
%
%
We find the expected behavior of the wave functions.
Interestingly, we find that a new peak appears near the naked singularity. 
This may be explained by the gravitational repulsion and the quantum effect. 
Namely, particles pushed outward by the strong repulsion are pushed back
toward the center by the exclusion principle.
This corresponds to the inverse of the phenomenon where the particles
localized at the center by gravitational attraction are pushed outward
by the exclusion principle.

\section{Black hole molecule} \label{sect:blackhole_molecule}
According to elementary chemistry, when two atomic nuclei share
electrons, a hydrogen molecule is formed.
In particular, the hydrogen molecular ion $\mathrm{H}_2^+$, where two atomic nuclei
share a single electron, is known as a model that can be solved
exactly.
It is now natural to examine the ``black hole molecule'' that consists
of two black hole hydrogen atoms.
This is composed of the two stable charged black holes together with the quantum
electrons that are trapped by them.
We consider the multicentered Majumdar--Papapetrou black hole solution as
the nuclei of the molecule. 
The Majumdar--Papapetrou solution in four dimensions is given by
\begin{align}
{\dop} s^2 
= - U^{-2} (\dop x^0)^2 + U^2 (\dop x^2 + \dop y^2 + \dop z^2), \qquad 
A = U^{-1} \, \dop x^0,
\label{eq:MP}
\end{align}
where $U$ is the harmonic function in $\mathbb{R}^3$, 
\begin{align}
U = 1 + \sum_{i=1}^N \frac{M_i}{R_i},
\qquad
R_i^2 = (x-x_i)^2 + (y-y_i)^2 + (z-z_i)^2.
\end{align}
Here, $M_i$ ($i=1, \ldots, N$) are constants and $(x_i, y_i, z_i)$ are
centers of the black holes in $\mathbb{R}^3$.
Each black hole in the Majumdar--Papapetrou solution \eqref{eq:MP} is stable since 
the gravitational attraction and the electromagnetic repulsion are
balanced.

\begin{figure}[t]
\centering
\includegraphics[scale=0.4]{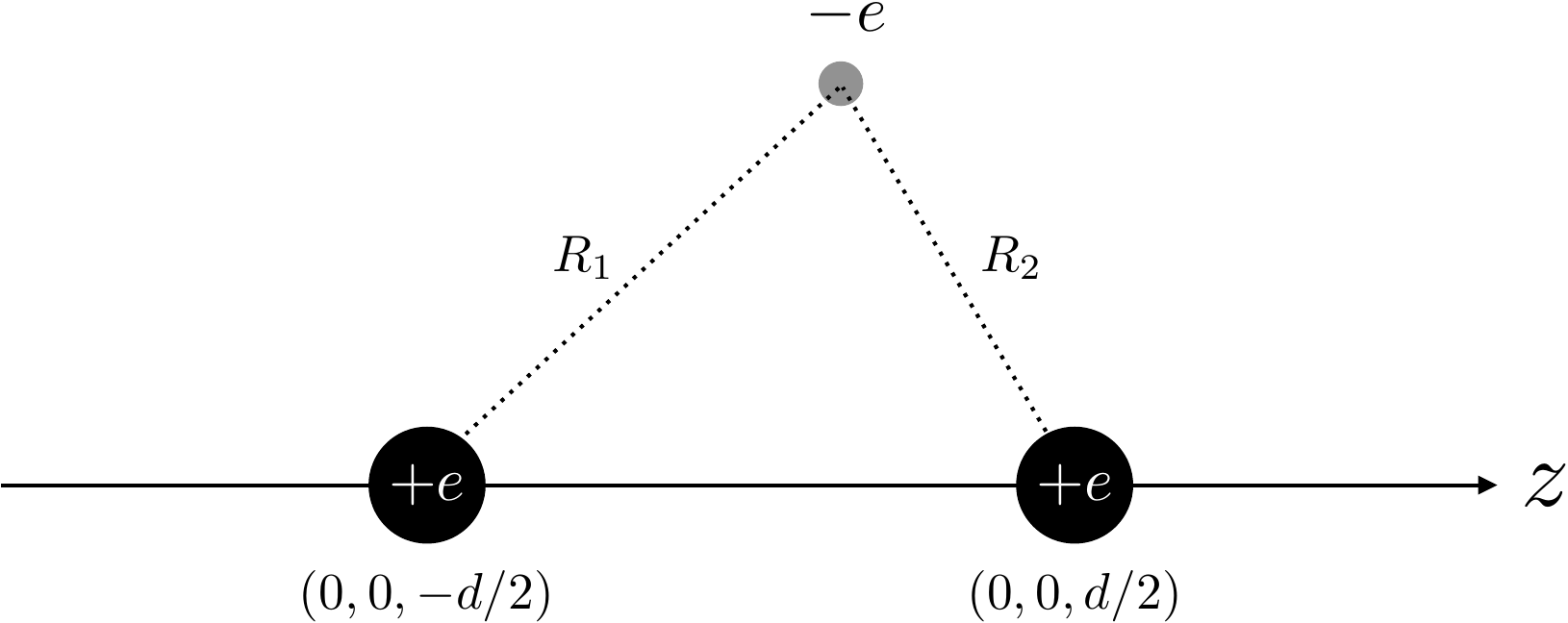}
\caption{%
A schematic picture of the hydrogen molecular ion $\mathrm{H}_2^+$ composed of two charged black holes and
 an electron.
}
\label{fig:H2}
\end{figure}

We now focus on the $N=2$ binary black holes as the prototypical example.
In this case, the three-dimensional Ricci scalar is calculated as
\begin{align}
R^{(3)} = 
\frac{
2 
\Big\{
M_2^2 R_1^4 + M_1^2 R_2^4 + 2 M_1 M_2 R_1 R_2 (\vec{R}_1 \cdot \vec{R}_2)
\Big\}
}
{
\Big(
M_2 R_1 + M_1 R_2 + R_1 R_2
\Big)^4
},
\end{align}
where 
$\vec{R}_i = (x - x_i, y - y_i, z - z_i)$ ($i=1,2$).
As in the case of the ordinary hydrogen molecule in the flat space, we employ the
Born--Oppenheimer approximation.
Namely, we ignore the dynamics of the nuclei (the binary black holes).
We also consider a single electron trapped on them for simplicity.
This corresponds to the hydrogen molecular ion $\mathrm{H}_2^+$ composed of the
black holes having each charge $+e$ and a single electron with mass $\mu$ and
the charge $-e$.
Then the Schr\"odinger equation is given by
\begin{align}
&
 -\frac{\hbar^{2}}{2 \mu}
\left\{
  \frac{1}{U^{2}}\,\nabla^{2}\psi
  + \frac{1}{U^{3}}\bigl(\nabla U \cdot \nabla \psi\bigr)
\right\}
- 
\left(
\frac{e^2}{R_1}
+
\frac{e^2}{R_2}
\right)
 \psi
\notag \\
& \qquad \qquad \qquad \qquad
+
\frac{\hbar^{2}}{12}\,
\frac{2\!\left\{
  M_{2}^{\,2} R_{1}^{4}
  + M_{1}^{\,2} R_{2}^{4}
  + 2 M_{1} M_{2} R_{1} R_{2}\,\bigl(\vec{R}_{1}\!\cdot\!\vec{R}_{2}\bigr)
\right\}}
{\bigl(M_{2} R_{1} + M_{1} R_{2} + R_{1} R_{2}\bigr)^{4}}
= E \psi.
\label{eq:MP_Schrodinger}
\end{align}
where we have included the potential $V (r) = - \frac{e^2}{R_1} -
\frac{e^2}{R_2}$ that provides the electromagnetic
attractive force.\footnote{
For the ordinary hydrogen molecular ion $\mathrm{H}_2^+$, 
 the potential includes $V (r) = \frac{e^2}{R_{12}}$ that
represents the repulsive force between two nuclei.
It is obvious that in our setup, such repulsive force does not exist.
}
For the standard $\mathrm{H}_2^+$ case, a variable separation is possible by
using the spheroidal coordinates.
We assume that the two black holes are located on the points $(0,0,-d/2)$
and $(0,0,d/2)$ in $\mathbb{R}^3$ and $M_1 = M_2 = M$ (Fig.~\ref{fig:H2}).

It is convenient to employ the coordinates
\begin{align}
\xi = \frac{R_1 + R_2}{d},
\qquad
\eta = \frac{R_1 - R_2}{d}.
\end{align}
These together with an azimuth angle $\phi$ consist of the spheroidal
coordinates.
They are defined in $1 \le \xi < \infty$, $-1 \le \eta \le 1$, and $0 \le
\phi < 2\pi$.
These are related to the Cartesian coordinate as 
\begin{align}
x = \frac{d}{2} \sqrt{(\xi^2 - 1) (1 - \eta^2)} \cos \phi,
\qquad
y = \frac{d}{2} \sqrt{(\xi^2 - 1) (1 - \eta^2)} \sin \phi,
\qquad
z = \frac{d}{2} \xi \eta.
\end{align}
By using this coordinate system, the Laplacian becomes
\begin{align}
\nabla^2 = \frac{4}{d^2 (\xi^2 - \eta^2)}
\Bigg[
\frac{\del}{\del \xi} 
\left(
(\xi^2 - 1) \frac{\del}{\del \xi}
\right)
+
\frac{\del}{\del \eta}
\left(
(1 - \eta^2) \frac{\del}{\del \eta}
\right)
+
\left(
\frac{1}{\xi^2 - 1}
+
\frac{1}{1 - \eta^2}
\right)
\frac{\del^2}{\del \phi^2}
\Bigg].
\end{align}
We also have 
\begin{align}
R^{(3)} = 
-
\frac{
128 M^2 (\xi^2 - \eta^2 - 3 \eta^2 \xi^2 - \xi^4)
}
{
\Big\{
4 M \xi - d (\xi^2 - \eta^2)
\Big\}^4
},
\qquad
U =
1 + \frac{4 M}{d} \frac{\xi}{\xi^2 - \eta^2}.
\end{align}
Then, using the decomposition $\psi (\xi, \eta, \phi) = R (\xi) S (\eta)
{\e}^{i \alpha \phi}$, the Schr\"odinger equation~\eqref{eq:MP_Schrodinger}
becomes
\begin{align}
& \scalebox{.96}{$
\begin{aligned} \displaystyle
&
\frac{1}{R} 
\frac{\dop}{\dop \xi}
\Big\{
(\xi^2 - 1) \frac{\dop}{\dop \xi} R
\Big\}
+
\frac{1}{S}
\frac{\dop}{\dop \eta}
\Big\{
(1 - \eta^2) \frac{\dop}{\dop \eta} S
\Big\}
-
\frac{\alpha^2}{\xi^2 - 1}
-
\frac{\alpha^2}{1 - \eta^2}
\notag \\
&
- \frac{1}{4} 
\frac{
4 M (\xi^2 + \eta^2) (\xi^2 - 1)
}
{
(\xi^2 - \eta^2)
\Big\{
d (\xi^2 - \eta^2) + 4 M \xi
\Big\}
}
\frac{1}{R} \frac{\dop R}{\dop \xi}
-
\frac{1}{4}
\frac{
8 M \eta \xi (1 - \eta^2)
}
{
(\xi^2 - \eta^2)
\Big\{
d (\xi^2 - \eta^2) + 4 M \xi
\Big\}
}
\frac{1}{S} \frac{\dop S}{\dop \eta}
\notag \\
& \ 
+ 
\left(
1 + \frac{4M}{d} \frac{\xi}{\xi^2 - \eta^2}
\right)^2
\left(
d A \xi 
+
\frac{2 \mu d^2}{3}
(\xi^2 - \eta^2)
\frac{
128 M^2 (\xi^2 - \eta^2 - 3 \eta^2 \xi^2 - \xi^4)
}
{
\Big\{
4 M \xi - d (\xi^2 - \eta^2)
\Big\}^4
}
-
\kappa^2 d^2 
(\xi^2 - \eta^2)
\right)
= 0,
\end{aligned}
$} \notag \\
& 
\label{eq:molecule}
\end{align}
where we have defined
$A = \frac{2 \mu e^2}{\hbar^2}$, $\kappa^2 = - \frac{2 \mu E}{\hbar^2}$, 
and used the following relation:
\begin{align}
&
\vec{\nabla} U \cdot \vec{\nabla} \psi 
=
\frac{1}{d^2}
\Bigg[
\frac{\xi^2 - 1}{\xi^2 - \eta^2}
\frac{\del U}{\del \xi} 
\frac{\del}{\del \xi} \psi
+
\frac{1 - \eta^2}{\xi^2 - \eta^2} \frac{\del U}{\del \eta} 
\frac{\del}{\del \eta} \psi
+
\frac{1}{(\xi^2 - 1) (1 - \eta^2)}
\frac{\del U}{\del \phi}
\frac{\del}{\del \phi}
\psi
\Bigg].
\end{align}
We find that the $\phi$ dependence is decoupled but the $\xi$ and $\eta$
sectors are not.
This is expected since the St\"ackel condition of the equation is
violated by the curvature corrections.
Therefore, the equation is solved by the numerical analysis.

In order to solve the equation analytically, it is necessary 
to decouple the two variables $\xi$ and $\eta$.
To this end, we consider a specific limit 
$\mu \to 0$ with fixed $M$.
This corresponds to the limit where the mass of the electron is small
compared to the mass of the black hole, and is a natural assumption.
Then the third line in \eqref{eq:molecule} is ignored, and the terms in the second line
in the large-$\xi$ region become
$
- \frac{M}{d} \frac{1}{R} \frac{\dop R}{\dop \xi}
$.
Then the equation in this asymptotic region becomes
\begin{align}
&
\frac{\dop}{\dop \xi}
\Big\{
(\xi^2 - 1) \frac{\dop}{\dop \xi} R
\Big\}
- \frac{M}{d} \frac{\dop R}{\dop \xi}
-
\Big(
\Lambda
+
\frac{\alpha^2}{\xi^2 - 1}
\Big) R = 0,
\notag \\
&
\frac{\dop}{\dop \eta}
\Big\{
(1 - \eta^2) \frac{\dop}{\dop \eta} S
\Big\}
+
\Big(
\Lambda
-
\frac{\alpha^2}{1 - \eta^2}
\Big) S = 0,
\end{align}
where $\Lambda$ is a parameter.
Now in this specific regime and parameters, the two equations are
decoupled and can be solved
analytically.
Note that the effects of the black holes $M \not= 0$ still remain in these equations.
We find solutions given by
\begin{align}
R(\xi) =& \ 
\left(
\frac{1-\xi}{1+\xi}
\right)^{\frac{M}{2d}}
\Big(
c_1 P^B_A (\xi)
+
c_2 Q^B_A (\xi)
\Big),
\notag \\
S(\eta) =& \ 
c_3 
P^{\alpha}_A
(\eta)
+
c_4 
Q^{\alpha}_A
(\eta),
\label{eq:H2}
\end{align}
where $P^B_A(x)$ and $Q^B_A (x)$ are the associated Legendre functions of 
the first and the second kinds, respectively.
The parameters are given by $A = \frac{1}{2} (\sqrt{4 \Lambda + 1} - 1)$ and $B = \frac{1}{2d}
\sqrt{M^2 + 4 d^2 \alpha^2}$, 
and $c_1, \ldots, c_4$ are integration constants.
The single-valuedness of the wave function requires that $\alpha$ be an
integer.
If we require the discrete energy spectra and the 
normalizability for bound states, $A$ and $B$ would be integers 
satisfying the conditions $0 \le B \le A$ and $0 \le \alpha \le A$.
It is known that the exact solution for the ordinary hydrogen molecule
ion $\mathrm{H}_2^+$ is given by 
the associated Legendre polynomials $P^m_n(x)$ 
and the confluent Heun function $H (x)$ \cite{Bates:1953}.
Although the solution \eqref{eq:H2} has a similar structure with the ordinary
$\mathrm{H}_2^+$ in the sense that they share 
the Legendre polynomials,
they are of course different due to the curvature 
corrections and the absence of the internuclei potential.

\section{Conclusion and discussions} \label{sect:conclusion}

In this paper, we studied quantum wave functions near black holes.
To this end, we utilized the formalism 
developed by DeWitt \cite{DeWitt:1957at}.
The equation itself has been known for a long time, 
but to our knowledge, this is the first time that it is applied to
black holes and its solutions are analyzed.
We in particular focused on the hydrogen atom type configuration, namely,
a single electron trapped by black holes.
We considered time slices of the Schwarzschild and Reissner--Nordstr\"om
black hole geometries and explicitly wrote down equations.

We first derived the equation for free particle in the spatial geometry of the 
Schwarzschild black hole.
We found that the equation is rewritten in the form of the confluent Heun
equation and it is, therefore, solved analytically.
In order to visualize the wave function, we found it convenient to solve
the equation numerically.
We showed that the wave function is attracted toward the black hole and
observed that particles with higher energy are more strongly attracted.
We then introduced a hypothetical electric attractive potential to the
Schwarzschild black hole and studied the bound states.
The equation is again rewritten in the form of the confluent Heun
equation.
We numerically found solutions satisfying appropriate boundary conditions
and visualized the wave functions for the hydrogen atom type
configuration.
We found that the wave function and the electron cloud are 
attracted into the horizon and are confined around the horizon.

Next, we focused on the charged black hole and discussed the genuine
black hole atom. 
We considered the Reissner--Nordstr\"om black hole solution and derived the
Schr\"odinger equation.
Assuming a spherically symmetric solution, we numerically solved the
equation in the radial direction. 
We then examined the effects of the black hole on the hydrogen atom for
various quantum numbers. 
We visualized the wave function, explicitly showing the difference
from the ordinary hydrogen atom. 
We observed that the electron cloud is generically bound to the outer
horizon.
We also investigated the wave function extending into the region inside
the inner horizon. 
We observed that the wave function is attracted to the inner horizon from
the interior. 
This is most naturally interpreted as the famous 
gravitational repulsion force present in the $r \sim 0$ region of the
Reissner--Nordstr\"om black hole.
These results indicate that not only classical objects but also quantum
wave functions are affected by gravity.
This is obvious intuitively (classically), but not evident in quantum
theory.
We also studied the wave functions in the extremal and the overextremal cases.

We then derived the Schr\"odinger equation for a molecule with a binary
black hole as its nucleus.
It is known that the Schr\"odinger equation for the hydrogen molecular
ion $\mathrm{H}_2^+$ in flat spacetime can be solved exactly.
We consider Majumdar--Papapetrou black holes as the nucleus.
Since there is no force acting between these black holes, 
it is in some sense simpler than ordinary molecules. 
However, due to the presence of the curvature term, the spheroidal
coordinates variables cannot be separated. 
Therefore, numerical analysis is generally required. 
We demonstrated that under specific conditions, the equation becomes
separable and analytical solutions can be obtained.

It would be helpful to know whether the black hole discussed in this paper is
available in the current Universe or not.
The Schwarzschild radius of a black hole of mass $M$ is given by 
$r_0 = \frac{2 G_{\mathrm{N}} M}{c^2}$, where $G_{\mathrm{N}}$ is Newton's gravitational
constant.
On the other hand, the Bohr radius of the hydrogen atom is known as 
$a_0 = \frac{4 \pi \varepsilon_0 \hbar^2}{m_e e^2} \sim 10^{-10} \, \mathrm{m}$ 
(where $m_e$ is the mass of the electron).
The mass of the black holes of the size of the hydrogen atom is estimated as 
$M = \frac{c^2 a_0}{2 G_{\mathrm{N}}} \sim 6.7 \times 10^{16} \, \mathrm{kg}$,
which is of the same order as the small asteroid.
Black holes of such a small mass are not created by the gravitational
collapse of stars but may appear in the early stage of the Universe.
The time $\Delta t$ that it takes for a black hole of mass $M \sim 10^{16} \, \mathrm{kg}$ 
to disappear through the Hawking radiation is estimated
as $\Delta t \sim \frac{G_{\mathrm{N}}}{\hbar c^3} M^3 \sim 10^{24} \, \mathrm{yr}$,
which is greater than the age of the Universe.
Therefore, it is possible that it remains in the current Universe as a primordial black hole.
It cannot be excluded that such black holes trap electrons and form the
``atoms'' discussed in this paper.

An interesting consideration related to what we discussed here is 
the discrepancy between the classical and quantum depictions of black
hole atoms.
In the purely classical picture, a test electron that
is not supported by any nongravitational force and has insufficient
angular momentum cannot remain at rest near a black hole.
In such a situation, the electron cannot help crossing the horizon in a
finite proper time along its timelike geodesic.
In contrast, in the ordinary hydrogen atom in the flat space, quantum
mechanics prevents the electron from collapsing into the nucleus and instead provides
stationary bound states. 
Our results suggest an analogous mechanism for the ``black hole atom,''
namely, quantum effects can support bound states of the electron 
even in regions where classical motion admits no stable circular orbit
so that the electron does not necessarily fall into the black hole.
The ``gravity versus quantum'' effects is an interesting issue to study.
It would be interesting to study this issue in more complex black holes, such
as Kerr solutions and so on.

\subsection*{Acknowledgments}
The work of S.~S. and K.~S. is supported by Grant-in-Aid for Scientific
Research, JSPS KAKENHI Grant Number JP25K07324.
The work of K.~S. is also supported by the Kitasato University Research Grant for Young Researchers. 


\end{document}